\documentclass[preprint,12pt]{elsarticle}
\usepackage{bm}
\usepackage{amssymb}

\makeatletter
\def\ps@headings{%
\def\@oddhead{\mbox{}\scriptsize\rightmark \hfil \thepage}%
\def\@evenhead{\scriptsize\thepage \hfil \leftmark\mbox{}}%
\def\@oddfoot{}%
\def\@evenfoot{}}
\makeatother
 \makeatletter
 \def\ps@headings{%
 \def\@oddhead{\mbox{}\scriptsize\rightmark \hfil \thepage}%
 \def\@evenhead{\scriptsize\thepage \hfil \leftmark\mbox{}}%
 \def\@oddfoot{}%
 \def\@evenfoot{}}
 \makeatother
 \usepackage{array}
 \usepackage[svgnames]{xcolor} 
 \usepackage{mdwmath}
 \usepackage{subfigure}
  \usepackage{bm}
 \usepackage{tipa}
  \usepackage{hyperref}
  \usepackage{framed}
  \usepackage{float}
 \usepackage{algorithm}
\usepackage{algorithmic}
\usepackage{overpic}
\usepackage{framed}
\usepackage{textcomp}
\usepackage{url}
\usepackage{graphicx}
\usepackage{times}
\usepackage{amsmath}
\PassOptionsToPackage{hyphens}{url}\usepackage{hyperref}
 \newtheorem{theorem}{Theorem}[section]
\newtheorem{lemma}[theorem]{Lemma}

 \newproof{proof}{Proof}
\journal{Computer Networks}

\begin{document}

\begin{frontmatter}



\title{Distributed Load Management Algorithms in Anycast-based CDNs\tnoteref{published}}
\tnotetext[published]{Part of the paper appeared in 53rd Annual Allerton Conference on Communication, Control and Computing – 2015.}

\author[sinha]{Abhishek Sinha\corref{cor1}}
\ead{sinhaa@mit.edu}
\address[sinha]{Massachusetts Institute of Technology, Cambridge, MA}
\author[micro]{Pradeepkumar Mani}
\ead{prmani@microsoft.com}
\author[micro_research]{Jie Liu}
\ead{Jie.Liu@microsoft.com}
\author[sales]{Ashley Flavel\corref{cor2}}
\ead{aflavel@salesforce.com}
\author[micro]{Dave Maltz}
\ead{dmaltz@microsoft.com}

\address[micro]{Microsoft Inc., Redmond, WA\fnref{micro}}
\address[sales]{Salesforce.com, Redmond, WA\fnref{sales}}
\address[micro_research]{Microsoft Research, Redmond, WA \fnref{micro_research}}
\cortext[cor1]{Corresponding Author}
\cortext[cor2]{This work was done when the author was an employee at Microsoft, Redmond.} 
\begin{abstract}
Anycast is an internet addressing protocol where multiple hosts share the same IP-address. A popular architecture for modern Content Distribution Networks (CDNs) for geo-replicated services consists of multiple layers of proxy nodes for service and co-located  DNS-servers for load-balancing among different proxies. Both the proxies and the DNS-servers use anycast addressing, which offers simplicity of design and high availability of service at the cost of partial loss of routing control. Due to the very nature of anycast, redirection actions by a DNS-server also affects loads at nearby proxies in the network. This makes the problem of optimal distributed load management highly challenging. In this paper, we propose and evaluate an analytical framework to formulate and solve the load-management problem in this context. We consider two distinct algorithms. In the first half of the paper, we pose the load-management problem as a convex optimization problem. Following a Kelly-type dual decomposition technique, we propose a fully-distributed load-management algorithm by introducing \emph{FastControl} packets. This algorithm utilizes the underlying anycast mechanism itself to enable effective coordination among the nodes, thus obviating the need for any external control channel. In the second half of the paper, we consider an alternative greedy load-management heuristic, currently in production in a major commercial CDN. We study its dynamical characteristics and analytically identify its operational and stability properties. Finally, we critically evaluate both the algorithms and explore their optimality-vs-complexity trade-off using trace-driven simulations.
\end{abstract}

\end{frontmatter}
\section{Introduction}
With the advent of ubiquitous computing and web-access, the last decade has witnessed an unprecedented growth of Internet-traffic. Popular websites, such as \texttt{bing.com}, registers more than a billion hits per day, which need to be processed efficiently in an online fashion, with as little latency as possible. 
Content Distribution Networks (CDN) are \emph{de facto} architectures to transparently reduce latency between the end-users and the web-services. In CDNs, a collection of non-origin servers (also known as \emph{proxies}) attempt to offload requests from the main servers located in the data-centers, by delivering cached contents on their behalf \cite{krishnamurthy2001use}.  Popular examples of internet services using CDN includes web objects, live-streaming media, emails and social networks \cite{saroiu2002analysis}. Edge-servers with cached contents serve as proxies to intercept some user requests and return contents without a round-trip to the data-centers. Online routing of user-requests to the optimal proxies remains a fundamental challenge in managing modern CDNs. Routing to a remote proxy may introduce extra round-trip delay, whereas routing to an overloaded proxy may cause the request to be dropped. 

Anycast is a relatively new paradigm for CDN-management and there are  already several commercial CDNs in place today using anycast \cite{NSDI_paper}, \cite{engel1998using}, \cite{swildens2009global}. With anycast, multiple proxy-servers, having the same user-content, share the same IP-address. Anycast relies on internet-routing protocols (such as, BGP) to route service-requests to any one of the geo-replicated proxies, over a \emph{cheap} network-path \cite{zaumen2000load}. Anycast based mechanisms have the advantage of being simple to deploy and maintain.  Being available as a service in IPv6 networks, no global topology or state information are required for its use \cite{basturk1997using}. 

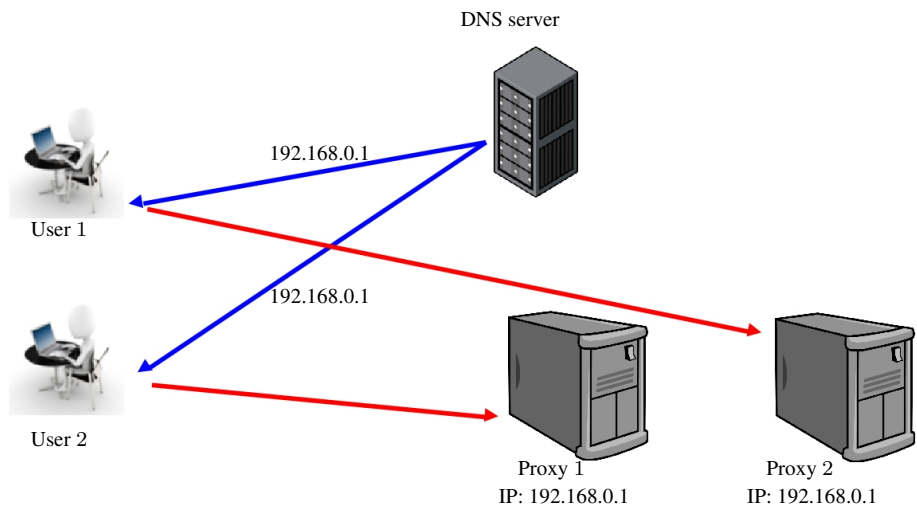
\begin{figure}
\center
\begin{overpic}[scale=0.3]{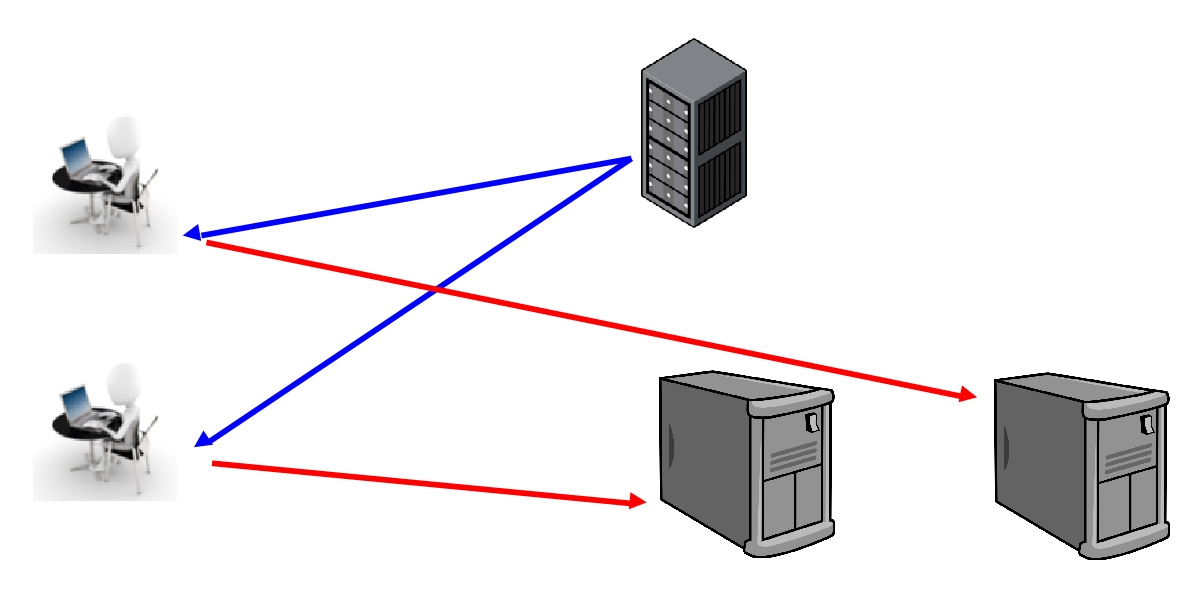}
\put(56,2){\scriptsize{Proxy $1$}} 
\put(54,-1){\scriptsize{IP: 192.168.0.1}}
\put(82,2){\scriptsize{Proxy $2$}} 
\put(80,-1){\scriptsize{IP: 192.168.0.1}}
\put(5,27){\scriptsize{User $1$}}
\put(5,5){\scriptsize{User $2$}}
\put(30,20){\scriptsize{192.168.0.1}}
\put(30,35){\scriptsize{192.168.0.1}}
\put(50,49){\scriptsize{DNS server}}
\end{overpic}
\caption{An example of Anycast-enabled CDN. User $1$ and $2$ obtain the anycast IP-address from the DNS server (blue-dotted arrow) and are then routed over the Internet \emph{either} to Proxy $1$ or to Proxy $2$ for service (red solid arrow). Note that both proxies have the same IP-address (Anycast addressing). The fraction of users landing on a given proxy is dictated by the Internet routing protocols and is beyond the control of CDN operators. This lack of load-awareness often leads to overloading of certain proxies.}
\label{anycast_ex}
\end{figure}

Although anycast routing can simplify the system-design and provide a high level of service-availability to the end-users \cite{sarat2006use}, it comes at the cost of partial loss of routing-control. This is because, a user-request may be routed to any one of the multiple geo-replicated proxies having the same anycast address. Since the user-to-proxy routing is done by the Internet routing protocols, which is not under the control of the CDN-operator, the user-request may end up in an already overloaded proxy, deteriorating its loading-condition further. Figure \ref{anycast_ex} illustrates the problem of load-management with Anycast.

 There have been several attempts in the literature to cope up with the lack of load-awareness issue with network-layer anycast. Papers \cite{miura2001server}, \cite{hashim2005active} consider ``Active Anycast" where additional intelligence is incorporated into the routers based on RTT and network congestion. It is specifically targeted to reduce pure latency rather than server overload,  thus yielding sub-optimal performance in a CDN setting. Alzoubi \emph{et al.} \cite{Alzoubi} formulates the anycast load-management problem as a General Assignment Problem, which is NP-hard. The paper \cite{jaseemuddin2006te} proposes a new CDN architecture which balances server-load and network-latency via detailed traffic engineering. 

\begin{figure}
\center
\begin{overpic}[scale=0.28]{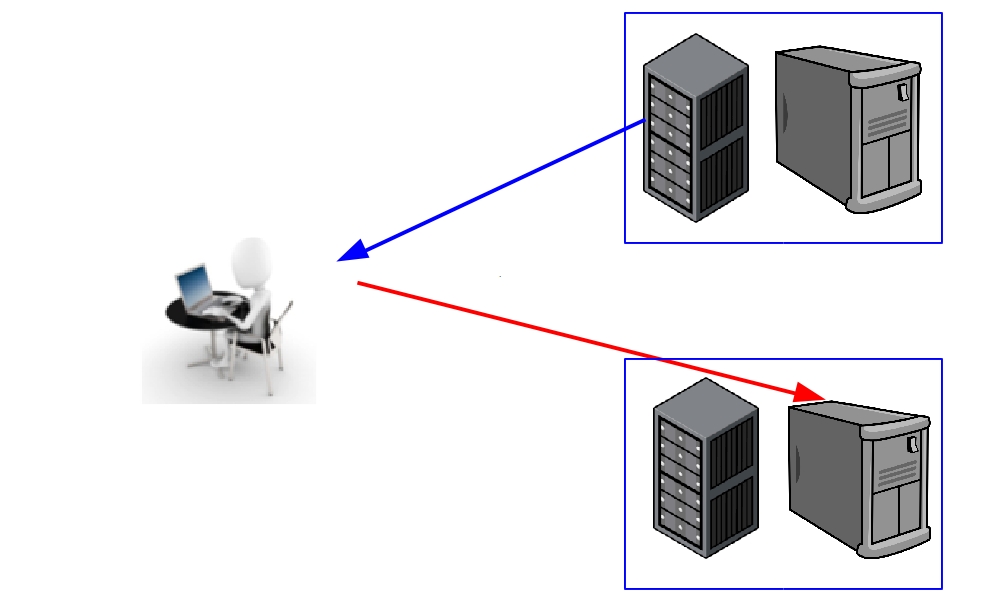}
\put(97,45){Node 1}
\put(64,65){\scriptsize{Co-located}}
\put(80,65){\scriptsize{Co-located}}
\put(67,62){\scriptsize{DNS}}
\put(83,62){\scriptsize{Proxy}}
\put(19,12){User}
\put(64,30){\scriptsize{Co-located}}
\put(80,30){\scriptsize{Co-located}}
\put(83,27){\scriptsize{Proxy}}
\put(67,27){\scriptsize{DNS}}
\put(42,33){$C_{12}$}
\put(97,10){Node 2}
\end{overpic}
\caption{Two \emph{FastRoute} CDN nodes with their corresponding co-located DNS and Proxy servers. Note that, although the user gets served by the proxy at Node $2$, it obtains the anycast IP-address from the DNS server at Node 1. Thus, in case of overload, this user can not be redirected from node $2$ by altering the DNS-response from the co-located DNS server at node $2$. }
\label{anycast_prob}
\end{figure}
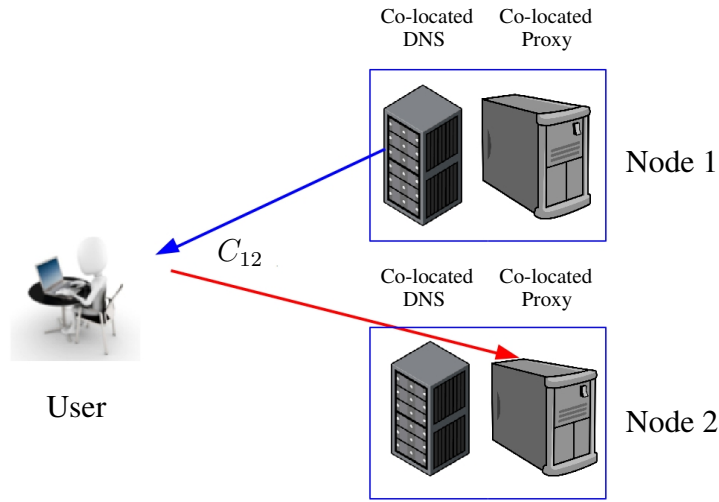

In this paper, we focus our attention to a state-of-the-art CDN-architecture, such as \emph{FastRoute} \cite{NSDI_paper}, which uses DNS-controlled-redirection for overload control in the proxies. Since DNS (Domain Name System) is the first point-of-contact of users to the internet, DNS-redirection is a popular and effective way to mitigate overload \cite{pang2004responsiveness}, \cite{shaikh2001effectiveness}. In this architecture, the proxies are logically arranged in layers of anycast rings, with each layer having a distinct Anycast IP-address. See Figure \ref{fig:layers} for an example. The provisioned capacities of the proxies increase as we move from the outer-layers to the inner-layers of the anycast rings. DNS is responsible for moving load across different layers by intelligently responding to users with different anycast addresses. Each proxy is equipped with a co-located authoritative DNS server. We will collectively refer to a proxy and its co-located DNS-server simply as a \emph{node}. Figure \ref{anycast_prob} and Figure \ref{fig:layers}  provide a high-level overview of the proposed CDN-architecture.
 
An \emph{overload} is said to occur when any proxy receives more service-requests than its processing-capacity. Since DNS is the primary control knob in this architecture, each DNS-server at a node is responsible for redirecting traffic to other layers to alleviate overload in the co-located proxy. A fundamental problem with this approach is that not all users, that hit a given proxy, can be redirected successfully by the co-located DNS. This is because, due to anycast routing, DNS-path and the data-flow paths are independent. Hence, an user's Local DNS (LDNS) could be obtaining a DNS-response from some authoritative DNS-server in a node, which is different from the DNS server co-located with the proxy which the user hits. An example is shown in Figure \ref{anycast_prob}. Hence, intuitively, the ability for a DNS-server at a node to control overload at the corresponding co-located proxy depends on the fraction of oncoming traffic to the proxy that are routed by the co-located DNS-server to the node itself. Informally, we refer to the above quantity as the \emph{self-correlation} \cite{NSDI_paper} of a given node. A formal definition of correlation in this context will be given in Section \ref{sec:model}.

Poor self-correlation could impair a node's ability to control overload in isolation. Hence, successful load management in layered CDN should involve coordinated action by DNS-servers in multiple nodes to alleviate overload. Thus the problem reduces to the DNS-plane determining the appropriate offload or redirection probabilities at each node to move traffic from the overloaded proxies to the next layer. This control-decision could be based on variety of information such as load on each proxy, DNS-HTTP correlation etc. From a practical point of view, not all of these quantities are easily measured and communicated to wherever is needed. Thus a centralized solution is not practically feasible and the challenge is to design a provably optimal, yet completely distributed load management algorithm.

Our key contributions in this paper are as follows:
\begin{itemize}
\item In section \ref{sec:model}, we present a simplified mathematical model for DNS-controlled load-management in modern anycast-based CDNs. Our model is general enough to address the essential operational problems faced by the CDN-operators yet tractable enough to draw meaningful analytical conclusions about its operational characteristics.

\item  In section \ref{optimization}, we formulate the load management problem as a convex optimization problem and propose a Kelly-type \emph{dual} algorithm \cite{kelly1997charging} to solve it in a distributed fashion. The key to our distributed implementation is the  \emph{Lagrangian decomposition} and the use of \emph{FastControl} packets, which exploits the underlying anycast architecture to enable coordination among the nodes in a distributed fashion. To the best of our knowledge, this is the first instance of such a decomposition technique employed in the context of load-management in CDNs.   

\item In section \ref{heuristic}, we consider an existing heuristic load-management algorithm used in \emph{FastRoute}, Microsoft\textquotesingle s CDN for many first party websites and services it owns \cite{azure}. We model the dynamics of this heuristic using non-linear system-theory and analytically derive its operational characteristics, which conform with the qualitative observations. To provide additional insight, a two-node system is analyzed in detail and it is shown, rather surprisingly, that given the ``self-correlations" of the nodes are sufficiently high, this heuristic load-management algorithm is able to control an incoming-load of \emph{any} magnitude, however large. Unfortunately, this theoretical guarantee breaks down once this correlation property no longer holds. In this case, the dual algorithm proposed in section \ref{optimization} performs better than the heuristic. 

	\item In section \ref{simulations}, we critically evaluate relative performance-benefits of the proposed optimal and the heuristic algorithm through extensive numerical simulations. Our simulation is trace-driven in the sense we use real correlation parameters collected over months from a large operational CDN \cite{NSDI_paper}. 
\end {itemize}

\section{System Model} \label{sec:model}

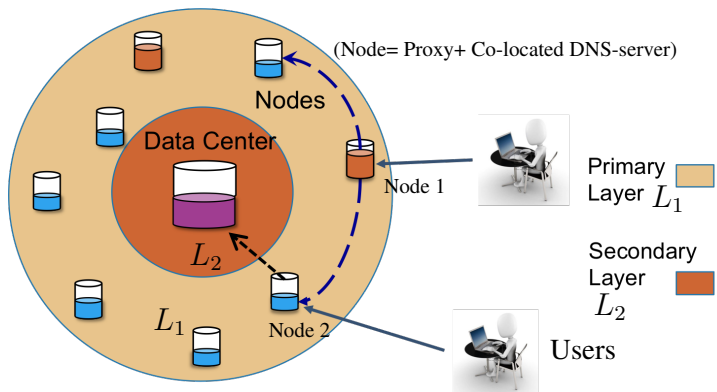
\begin{figure}[]
\center
\begin{overpic}[width=0.7\textwidth]{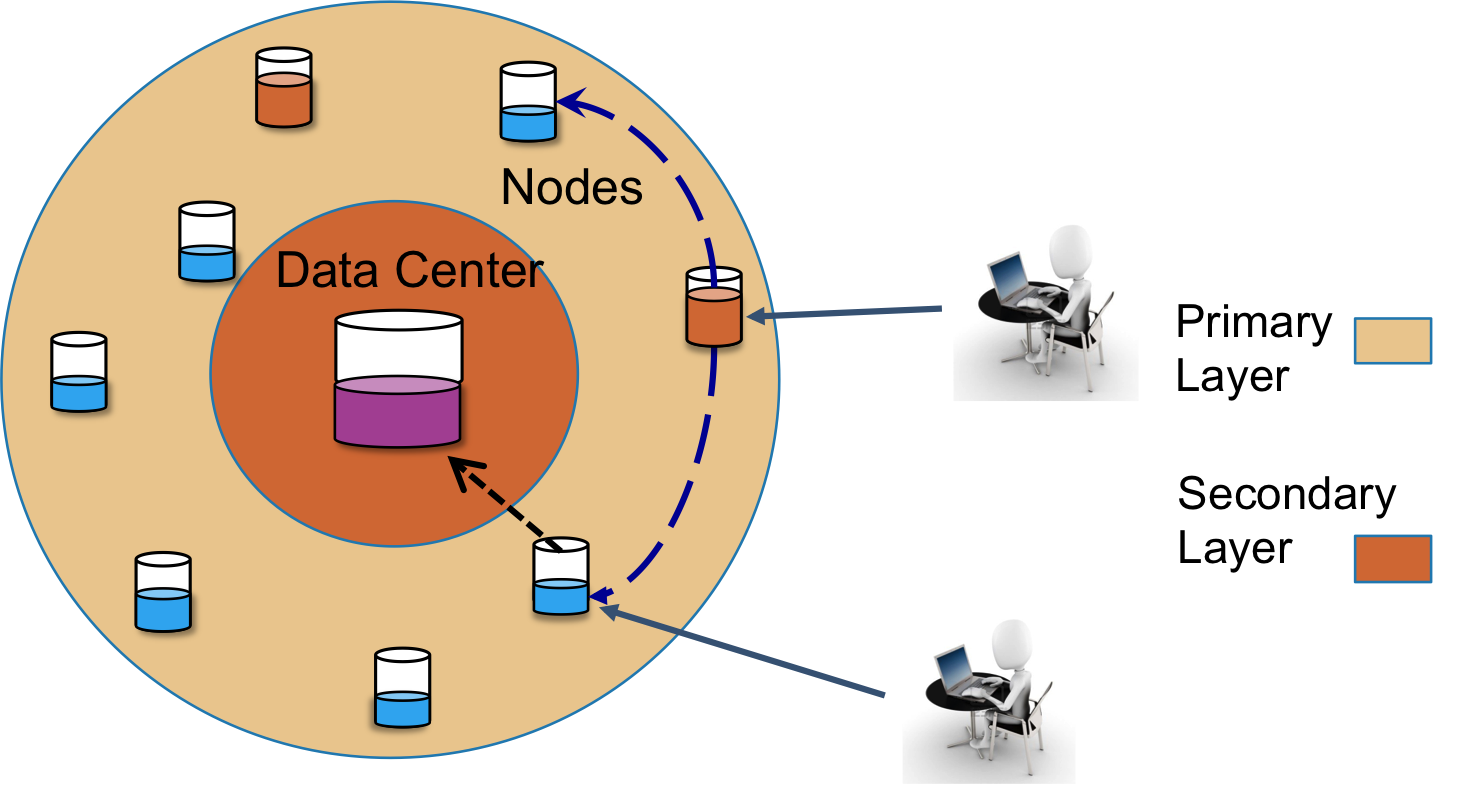}
\put(89,26){$\scriptsize{L_1}$}
\put(20,9){{$\small{L_1}$ }}
\put(52,28){\scriptsize{Node 1}}
\put(36,8){\scriptsize{Node 2}}
\put(25,18){{$L_2$}}
\put(81,11){$L_2$}
\put(45,47){\scriptsize{(Node= Proxy+ Co-located DNS-server)}}
\put(75,5){\small{Users}}
\end{overpic}
\caption{A CDN architecture with two anycast layers.  Solid arrows denote user's DNS path. Node $1$'s DNS returns the anycast address for Layer $1$, so, due to anycast, the user gets redirected to other nodes in $L_1$. Node $2$'s DNS returns the anycast address for $L_2$. So the user is redirected to the data-center. } 
\label{fig:layers}
\end{figure}
\indent \textbf{Nodes and Layers:} We consider an anycast-based CDN system consisting of two logical anycast layers: \emph{primary} (also referred to as $L_1$) and \emph{secondary} (also referred to as $L_2$). See Figure \ref{fig:layers} for a graphic depiction. The primary layer ($L_1$) hosts a total of $N$ nodes. Each node consists of a co-located authoritative DNS and a proxy server. See Figure \ref{anycast_prob} for a schematic diagram. The proxy servers are the end-points for user-requests (e.g. HTTP). The $i$\textsuperscript{th} proxy has a (finite) processing-capacity of $T_i$ user-requests per unit time. The secondary layer ($L_2$) consists of a single data-center, with practically infinite processing-capacity. Since the proxies in $L_1$ are geographically distributed throughout the world, an average user is typically located near to at least one proxy, resulting in relatively small user-to-proxy round-trip latency. On the other hand, the single data-center in $L_2$ is typically located further from the average user-base, resulting in significantly high user-to-data-center round-trip latency. Total response time for a user is the sum of round-trip latency and processing time of the server (either proxy or the data-center), which we would like to minimize. \\
\indent \emph{Anycast Addressing :} As discussed earlier, \emph{all} proxies in the primary layer share the same IP-address $\mathcal{I}_1$ and the data-center in $L_2$ has a distinct IP-address $\mathcal{I}_2$. This method of multiple proxies sharing the same IP-addresses is known as \emph{anycast} addressing and is widely used for geo-replicated cloud services \cite{cloudflareanycast}, \cite{Alzoubi}.\\
\indent \textbf{Control actions :} When a DNS-request by an user \footnote{In the Internet DNS-requests are actually generated by the Local DNS (LDNS) of the user and are recursively routed to an authoritative DNS server. However, to keep the model and the analysis simple, we will assume that individual DNS-queries are submitted by the users themselves. } arrives at a node $i$ in $L_1$, requesting for the IP-address of a server, the co-located DNS-server takes \emph{one} of the following actions
\begin{itemize}
\item \textbf{(1)} It either returns the address $\mathcal{I}_2$ (which offloads the user-request to the data-center) or,
\item \textbf{(2)} it returns the anycast address $\mathcal{I}_1$ (which, instead, serves the user-request in some proxy in the primary layer $L_1$ itself).
\end{itemize}
 This (potentially randomized) \textbf{binary decision} by the co-located DNS-server of node $i$ could be based on the following local variables measurable at node $i$ : 
\begin{enumerate}
 \item \textbf{\emph{DNS-influenced request} arrival rate at node $i$'s co-located DNS-server}, denoted by $A_i$ requests per unit time\footnote{We assume that $A_i$'s are piecewise constant and do not change considerably during the transient period of the load-management algorithms discussed here.}. Each DNS-request accounts for a certain amount of user-generated load which is routed to $L_1$ or $L_2$ according to the DNS-response by node $i$; We normalize $A_i$ so that each unit of $A_i$ corresponds to a unit of user-load; As an example, if 10\% of DNS responses at node $i$ returns the address $\mathcal{I}_2$, then the total amount of load shifted to the data-center ($L_2$) due to node $i$'s DNS is $0.1A_i$.
 \item \textbf{\emph{User-load} arrival rate at node $i$'s co-located proxy}, denoted by $S_i(t)$ requests per unit time. This is clearly a function of the DNS-arrivals $A_i$'s and the redirection-decisions of the co-located DNS of different nodes (See Eqn.\eqref{S_eqn}). The variable $S_i(t)$ is available locally at node $i$ (e.g., HTTP connection-request arrival-rates at the co-located proxy-server, in case of web-applications). We assume that the workloads induced by different user-requests have very small variability (which is true for web search query traffic, e.g.).

\end{enumerate}

\indent \textbf{Anycasting and inter-node coupling:} When a DNS-request arrives at a node $i$ and the co-located DNS-server returns the $L_1$ anycast-address $\mathcal{I}_1$, the corresponding user-request may be routed to any of one of the $N$ proxy-servers (all having the same IP-address $\mathcal{I}_1$) in the primary layer for service. The particular proxy, where the user-request actually gets routed to, depends on the corresponding ISP's routing policy, network congestion and many other factors. We assume that a typical DNS-query, which is routed to $L_1$ by the node $i$, lands at node $j$'s proxy for service with probability $C_{ij}$, where
\begin{eqnarray} \label{corr_mat_prop}
 \sum_{j=1}^{N}C_{ij}=1, \hspace{10pt}\forall i=1,2,\ldots, N
\end{eqnarray}
 In our system, we have determined the matrix $\bm{C}\equiv [C_{ij}]$ empirically by setting up an experiment similar to the one described in \cite{shaikh2001effectiveness}, using data collected over several months.\\
 \indent  In the ideal case, we would like to have $\bm{C} \approx \bm{I}$, where $\bm{I}$ is the $N \times N$ identity matrix. In this case, the nodes operate in an isolated fashion and situations like Figure \ref{anycast_prob} rarely takes place. However, as reported in \cite{NSDI_paper}, in real CDN-systems with ever-increasing number of proxies, the node are often highly-correlated. In this paper, we investigate the consequences that arise from non-trivial correlations among different nodes and its implications for designing load-management algorithms for CDNs. 

\indent \textbf{System Equations:} Assume that, due to action of some control-strategy $\pi$, the co-located DNS at node $i$ decides to randomly redirect $1-x_i^\pi(t)$ fraction of incoming DNS-queries (given by $A_i$) to Layer $L_2$ at time $t$ ($0\leq x_i^\pi(t) \leq 1$). Thus it routes $x_i^\pi(t)$ fraction of the incoming requests to different proxies in the layer $L_1$. Hence the total user-load arrival rate, $S_i(t)$, at node $i$'s co-located proxy may be written as 
\begin{eqnarray} \label{S_eqn}
 S_i(t)=\sum_{j=1}^{N} C_{ji}A_jx_j^\pi(t), \hspace{10pt}\forall i=1,2,\ldots, N
\end{eqnarray}
A local control strategy $\pi$ is identified by a collection of mappings $\bm{\pi}=\bigg(x^\pi_i(\cdot),i=1,2,\ldots,N\bigg)$, given by $\bm{x}^\pi_i: \Omega^t_i \times t \to [0,1]$, where $\Omega^t_i$ is the set of all observables at node $i$ up to time $t$.
\section{An Optimization Framework} \label{optimization}
\subsection{Motivation} 
The central objective of a load-management policy $\pi$ in a two-layer Anycast-CDN is to route as few requests as possible to the Data-Center $L_2$ (due to its high round-trip latency), without overloading the primary-layer $L_1$ proxies (due to their limited capacities). Clearly, these two objectives are at conflict with each other and we need to find a suitable compromise. The added difficulty, which makes the problem fundamentally challenging is that, each node is an autonomous agent and takes its redirection decisions on its own, based on its \textbf{local observables only}. As an example, a simple locally-greedy heuristic for node $i$ would be to redirect requests to $L_2$ (i.e. decrease $x_i(t)$) whenever its co-located proxy is overloaded (i.e., $S_i(t)>T_i$) and redirect requests to $L_1$ (i.e., increase $x_i(t)$) whenever the co-located proxy is under-loaded (i.e., $S_i(t)<T_i$). This policy forms the basis of the greedy control-strategy used in \cite{NSDI_paper}. \\
 This greedy strategy appears to be quite appealing for deployment, due to its extreme simplicity. However, in the next subsection \ref{u_o}, we show by a simple example that, in the presence of significantly high cross-correlations among the nodes (i.e. non-negligible value of $C_{ij}, i\neq j$), this simple heuristic could lead to an  \emph{uncontrollable overload situation}, an extremely inefficient operating point with degraded service quality. This example will serve as a motivation to come up with a more efficient distributed load-management algorithm, that we develop subsequently. \\

\subsection{Locally Uncontrollable Overload: An Example} \label{u_o}
\begin{figure}
\small
 \centering 
  \begin{overpic}[scale=0.9]{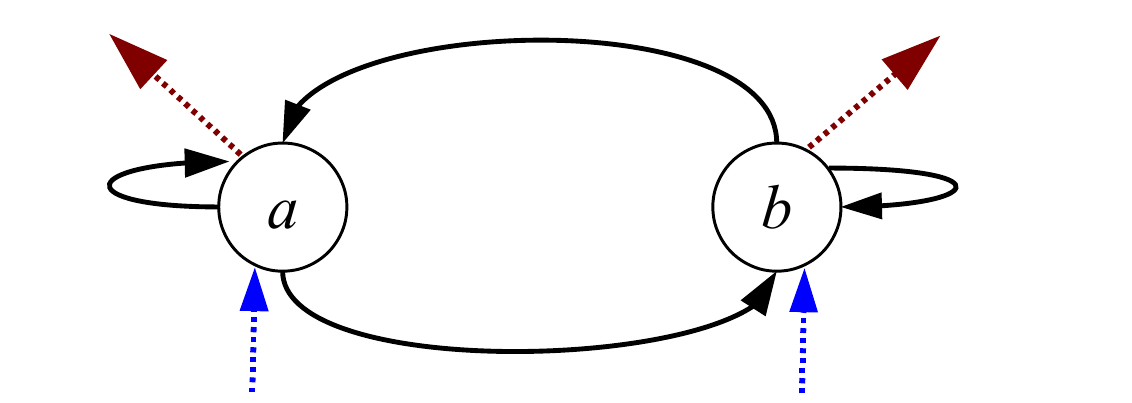}
  \put(10,13){$C_{aa}$}
  \put(8,9){$=0.1$}
  \put(38,-1){$C_{ab}=0.9$}
  \put(38,35){$C_{ba}=0.5$}
  \put(74,12){$C_{bb}=0.5$}
  \put(18,-2){$A_a=1$}
  \put(67,-2){$A_b=1$}
  \put(30,21){$T_a=0.7$}
  \put(32,15){$x_a(t)$}
  \put(53,15){$x_b(t)$}
  \put(51,21){$T_b=0.7$}
  \put(45,7){$L_1$}
  \put(4,43){Traffic to}
  \put(9,39){$L_2$}
  \put(5,35){$A_a(1-x_a(t))$}
  \put(71,43){Traffic to}
  \put(76,39){$L_2$}
  \put(72,35){$A_b(1-x_b(t))$}
   \end{overpic}
\caption{A two-node system illustrating locally uncontrollable overload at the node $b$}
\label{two-node_fig}
\end{figure}
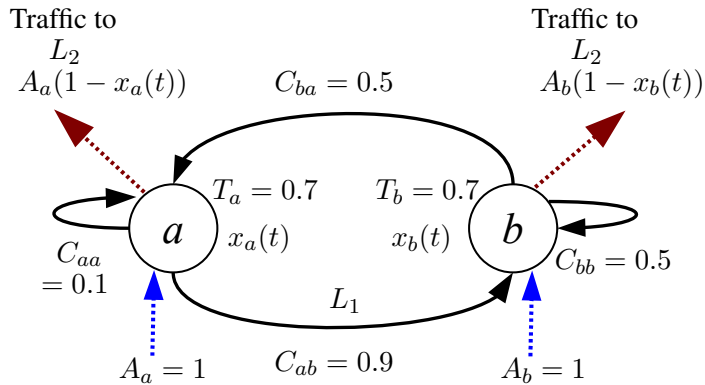

 Consider a two-layered CDN, hosting only two nodes $a$ and $b$ in the primary layer $L_1$, as shown in Figure \ref{two-node_fig}. The (normalized) DNS-request arrival rates to the nodes $a$ and $b$ are  $A_a=1$ and $A_b=1$. Suppose the processing capacities (also referred to as \emph{thresholds}) of the co-located proxies are $T_a=0.7$ and $T_b=0.7$ respectively. With the correlation components $[C_{ij}]$ as shown in Figure \ref{two-node_fig}, the user-loads at the co-located proxies are given as
 \begin{eqnarray}
  S_a(t)=0.1x_a(t)+0.5x_b(t)\\
  S_b(t)=0.9x_a(t)+0.5x_b(t)
 \end{eqnarray}
Since $0\leq x_a(t), x_b(t)\leq 1$, it is clear that under any control policy $\bm{x}(t)$ the following holds
\begin{eqnarray*}
	S_a(t) \leq 0.1\times 1 + 0.5 \times 1=0.6<0.7=T_a, \hspace{5pt} \forall t
\end{eqnarray*}
 Thus, the proxy at node $a$ will be under-loaded irrespective of the load-management policy $\pi$ in use. Consequently, under the greedy-heuristic (formally, Algorithm \ref{algo_greedy} in Section \ref{heuristic}), the co-located DNS-server at node $a$ will \emph{greedily} increase its $L_1$ redirection probability $x_a(t)$ such that $x_a(t) \nearrow 1$ in the steady-state (note that, node $a$ acts autonomously as it does not have node $b$'s loading information). This, in turn, overloads the co-located proxy in node $b$ because the steady-state user-load at proxy $b$ becomes 
\begin{eqnarray*}
	S_b(\infty) &=& 0.9 x_a(\infty)+0.5x_b(\infty)\\
	&=& 0.9\times 1 + 0.5x_b(\infty)\\
	&\geq& 0.9 >0.7=T_b.
\end{eqnarray*}
 Since node $b$ is overloaded in the steady-state, under the action of the above greedy heuristic, it will (unsuccessfully) try to avoid the overload by offloading the incoming DNS-queries to $L_2$, as much as it can, by letting $x_b(t)\searrow 0$. Thus the steady-state operating point of the algorithm will be $x_a(\infty)=1,x_b(\infty)=0$, with node $b$ overloaded. It is interesting to note that poor self-correlation of node $a$ ($C_{aa}=0.1$) causes the other node $b$ to overload, even under the symmetric DNS-request arrival patterns. Also, this conclusion does not depend on the detailed control-dynamics of the offload probabilities (viz. the instantaneous values if $\dot{x}_1(t)$ and $\dot{x}_2(t)$). Since the overload condition at the node-$b$ can not be overcome by isolated autonomous action of node-$b$ itself, we say that node-$b$ is undergoing a \emph{locally uncontrollable overload situation}. \\
 From the CDN-system point-of-view, the above situation is an extremely inefficient operating-point, because a large fraction ($45\%$ in the above example) of the incoming user-requests either gets dropped or severely-delayed due to the overloaded node-$b$. This poor operating point could have been potentially avoided provided the nodes somehow mutually co-ordinate their actions\footnote{Another trivial solution to avoid overload could be to offload all traffic from all nodes to $L_2$, i.e. $x_i(t)=0, \forall i, \forall t$. However, this is highly inefficient because it is tantamount to not using the proxies in the Primary layer $L_1$ at all.}. 
It is not difficult to realize that the principal reasons behind the locally uncontrollable overload situation in the above example are as follows:
 \begin{itemize}
  \item (1) distributed control with local information 
  \item (2) poor self-correlation of node $a$ ($C_{aa}=0.1$)
 \end{itemize}
The factor (1) is fundamentally related to the distributed nature of the system and requires coordinations among the nodes. In our distributed algorithm [\ref{algo_dual}] we address this issue by introducing the novel idea of \emph{FastControl} packets. This strategy does not require any explicit state or control-information exchange.\\
Regarding the factor (2), we intuitively expect that the local greedy-heuristic should work well if the self-correlation of the nodes (i.e. $C_{ii}$) are not too small. In this favorable case, the system will be loosely coupled, so that each node accounts for a major portion of the oncoming load to itself. In section \ref{heuristic}, we will return to a variant of this local heuristic used in FastRoute \cite{NSDI_paper} and derive analytical conditions under which the above intuition holds good.\\
In this section, we take a principled approach and propose an iterative load-management algorithm, which is provably optimal for arbitrary system-parameters ($\bm{A},\bm{C}$). In this algorithm, it is enough for each node $i$ to know its own local DNS and user-load arrival rates (i.e., $A_i$ and $S_i(t)$ respectively) and the entries corresponding to the $i$\textsuperscript{th} row and column of the correlation matrix $\bm{C}$ (i.e., $C_{i\cdot}, C_{\cdot i}$). No non-local knowledge of the dynamic loading conditions of other nodes $j\neq i$ is required for its operation. 

\subsection{Mathematical formulation}
\begin{figure}
\center
   \hspace{-60pt}
    \begin{overpic}[scale=0.28]{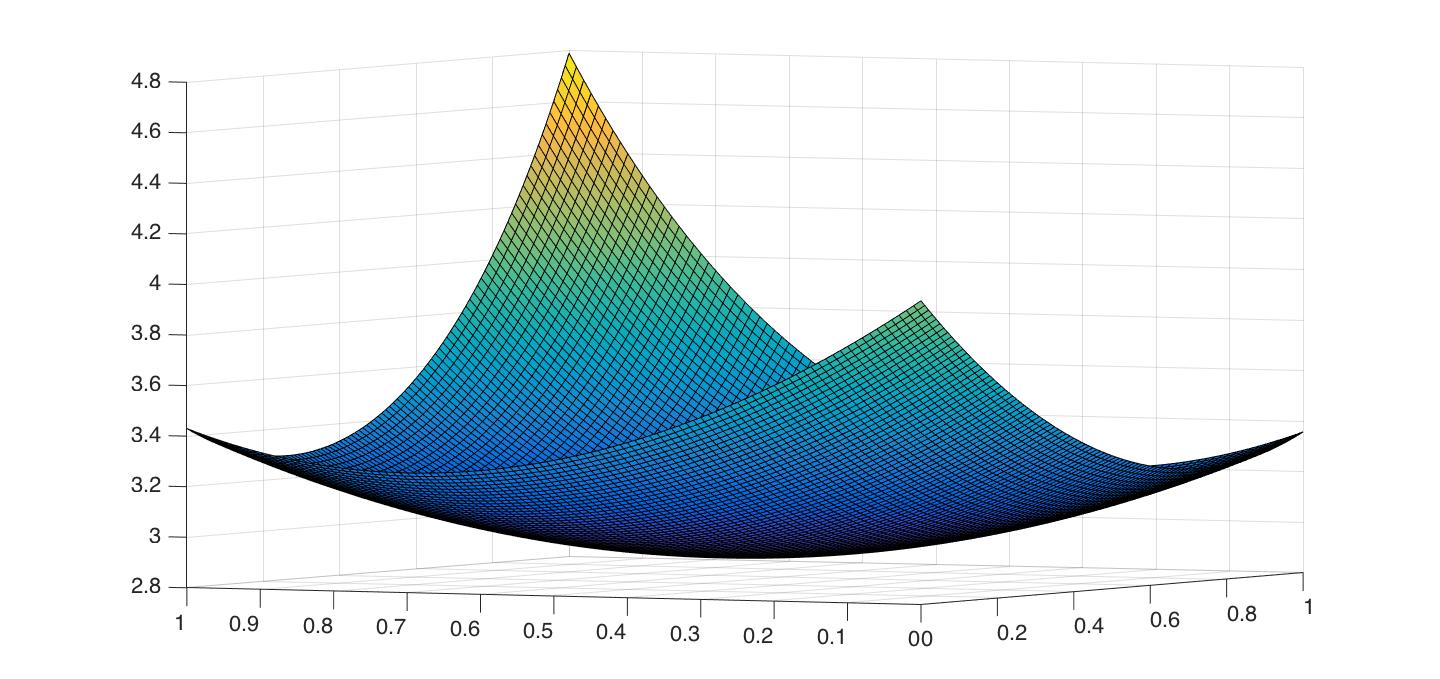}	
    \put(80,1){$x_1$}
    \put(40,1){$x_2$}
  \end{overpic}
  \caption{A typical plot of cost-surface for a two-node system as a function of their redirection probabilities $x_1,x_2 (0\leq x_1,x_2 \leq 1)$.}
  \label{cost_functions}
\end{figure}

We consider the following optimization problem. The variables $\bm{x}$ and  $\bm{S}$ have the same interpretation as above. The cost-functions, constraints and their connection to the load-management problem are discussed in detail subsequently.
\begin{framed}
\begin{eqnarray}\label{obj_fun}
 \text{Minimize} \hspace{20pt} W(\bm{x},\bm{S})\equiv \sum_{i=1}^{N}\big(g_i(S_i) + h_i(x_i)\big) 
\end{eqnarray}
Subject to,
\begin{eqnarray} \label{load_x}
S_i=\sum_{j=1}^{N}C_{ji}A_jx_j,\hspace{10pt} \forall i=1,2,\ldots, N
\end{eqnarray}
\begin{eqnarray*}
\bm{x}\in X, 
\bm{S} \in \Sigma_T
\end{eqnarray*}
\end{framed}
 \paragraph*{Discussion}The \underline{first component} of the cost function, $g_i(S_i)$, denotes the cost for serving $S_i$ amount of user-requests by the $i$\textsuperscript{th} proxy in $L_1$ per unit time. Clearly, this cost-component grows rapidly once the proxy is nearly over-loaded, i.e. $S_i \approx T_i$. In our numerical work, we take $g_i(\cdot)$ to be proportional to the average aggregate queuing delay for an $M/G/1$ queue with server-capacity $T_i$ \cite{bertsekas1987data}, i.e., 
\begin{eqnarray} \label{ex1}
g_i(S_i)&=& \begin{cases}
\frac{\eta_i S_i}{1-\frac{S_i}{T_i}}, \hspace*{20pt}\text{if   } S_i\leq T_i \\
 \infty, \hspace*{10pt} \text{otherwise} 
\end{cases}
\end{eqnarray}
Here $\eta_i$ is a positive constant, denoting the relative cost per unit increment in latency.\\
The \underline{second component} of the cost function $h_i(x_i)$ denotes the cost due to \emph{round-trip-latency} of requests routed to the Data-center ($L_2$). As an example, in a popular model \cite{roughgarden2005selfish} the delay incurred by a single packet over a congested path varies \emph{affinely} with the offered load. Since the rate of traffic sent to the secondary layer by node $i$ is $A_i(1-x_i)$, according to this model, the cost-function $h_i(x_i)$ may be taken as follows  
\begin{eqnarray} \label{ex2}
h_i(x_i)=\theta_i A_i(1-x_i)\big(d_i+\gamma_i A_i(1-x_i)\big)
\end{eqnarray}
where $d_i$ is a (suitably normalized) round-trip-latency parameter from the node $i$ to the data-center in $L_2$ and $\theta_i,\gamma_i$ are suitable positive constants. We use the cost-functions in Eqns. \eqref{ex1} and \eqref{ex2} for our numerical work. A typical plot of the cost-surface for the case of a two-node system as in Figure \ref{two-node_fig} is shown in Figure \ref{cost_functions}. \\

The \underline{constraint set} $X=[0,1]^N$ represents the $N$-dimensional unit hypercube in which the (controlled) redirection probabilities must lie and the set $\Sigma_T$ captures the capacity constraints of the proxies which are downward closed, e.g., if the proxy $i$ has capacity $T_i$ then we have  
\begin{eqnarray*}
 \Sigma_{\bm{T}}=\{\bm{S}: S_i \leq T_i, \forall i=1,2,\ldots, N\}
\end{eqnarray*}
The functions $g_i(\cdot), h_i(\cdot)$ are assumed to be closed, proper and strictly convex \cite{bertsekas1999nonlinear}. We also assume the functions $g_i(\cdot)$ to be monotonically increasing. Hence, we can replace the equality constraint \eqref{load_x} by the following inequality constraint, without loss of optimality
\begin{eqnarray*}
\sum_{j=1}^{N}C_{ji}A_jx_j \leq S_i,\hspace{10pt} \forall i=1,2,\ldots, N
\end{eqnarray*} 
This is because, if the optimal $S_i^*$ is strictly greater than the LHS, we can strictly (and feasibly) reduce the objective value by reducing $S_i^*$ to the level of LHS, resulting in contradiction.\\
Hence, the above load management problem is equivalent to the following optimization problem $\textbf{P}_1$ :\\
\begin{framed}
\textbf{Minimize} 
\hspace{10pt} $W(\bm{x},\bm{S})= \sum_{i=1}^{N}\big(g_i(S_i) + h_i(x_i)\big) $\\
\textbf{Subject to},
\begin{eqnarray}
&&\sum_{j=1}^{N}C_{ji}A_jx_j \leq S_i,\hspace{10pt} \forall i=1,2,\ldots, N \label{constr1}\\
&&\bm{x}\in X, \bm{S} \in \Sigma_{\bm{T}},
\end{eqnarray} 
where, $X=[0,1]^N$ and $\Sigma_{\bm{T}}=\{\bm{S}: S_i \leq T_i, \forall i=1,2,\ldots, N\}$. 
\end{framed}

Since with the above assumptions, the objective function and the constraint sets of the problem $\bm{P}_1$ are all convex \cite{bertsekas1999nonlinear}, we immediately have the following lemma:
\begin{framed}
\begin{lemma}
 The problem $\textbf{P}_1$ is convex. 
\end{lemma}
\end{framed}
Hence a host of methods \cite{bertsekas2015convex} can be used to solve the problem $\textbf{P}_1$ in a centralized fashion. However, we are interested in finding a load balancing algorithm for each node, which collectively solve the problem $\textbf{P}_1$ with locally available loading information only. This problem is explored in the next subsection. 

\subsection{The Dual Decomposition Algorithm } 
\label{dual_section}
In this section we derive a dual algorithm \cite{kelly1997charging}, \cite{lobel2011distributed}, \cite{eryilmaz2010distributed} for the problem $\textbf{P}_1$ and show how it leads to a distributed implementation with negligible control overhead.   \\
By associating a non-negative dual variable $\mu_i$ to the $i$\textsuperscript{th} constraint in \eqref{constr1} for all $i$, the Lagrangian of $\textbf{P}_1$ may be written as follows:

\begin{eqnarray}\label{dual_obj}
\mathcal{L}(\bm{x}, \bm{S}, \bm{\mu})= \sum_{i=1}^{N}\big( g_i(S_i)-\mu_iS_i) \big)+
\sum_{i=1}^{N}\bigg(h_i(x_i)+A_ix_i\big(\sum_{j=1}^{N}\mu_jC_{ij}\big)\bigg)
\end{eqnarray}
This leads to the following dual objective function \cite{bertsekas1999nonlinear}
\begin{eqnarray} \label{dual_opt}
D(\bm{\mu})=\inf_{\bm{x}\in X,\bm{S}\in \Sigma_{\bm{T}}} \mathcal{L}(\bm{x}, \bm{S}, \bm{\mu})
\end{eqnarray}
We now exploit the \emph{separability} property of the dual objective \eqref{dual_obj} to reduce the problem \eqref{dual_opt} into following two one-dimensional sub-problems: 
\begin{equation} \label{dual_sep}
\left.
\begin{aligned} 
S_i^*(\bm{\mu})&=\inf_{0\leq S_i\leq T_i} \bigg(g_i(S_i)-\mu_iS_i  \bigg) \quad \\
x_i^*(\bm{\mu})&=\inf_{0\leq x_i\leq 1}\bigg(h_i(x_i)+A_i\beta_i(\bm{\mu})x_i \bigg),
\end{aligned}
\right\}
\end{equation}
where the scalar $\beta_i(\bm{\mu})$ is defined as the (scaled) linear projection of the dual-vector $\bm{\mu}$ on the $i$\textsuperscript{th} row of the correlation matrix, i.e.
\begin{eqnarray} \label{beta_eqn}
\beta_i(\bm{\mu})=\sum_{j=1}^{N}\mu_jC_{ij} =\bm{C}_i^{T}\bm{\mu},
\end{eqnarray}
and $\bm{C}_i$ is the $i$\textsuperscript{th} row of the correlation-matrix $\bm{C}$.
\paragraph{Discussion} The scalar $\beta_i(\bm{\mu})$ couples the offload decision of node $i$ with the state of the entire network through Eqn.~\eqref{dual_sep}. Once the value of $\beta_i(\bm{\mu})$ is available to the node $i$, the node has all the required information at its disposal to \emph{locally} solve the corresponding sub-problems \eqref{dual_sep}, for a fixed value of $\bm{\mu}$. The solutions of these one-dimensional sub-problems may even be obtained in closed form in some cases. Also, note that $\beta_i(\bm{\mu})$, being a scalar, is potentially easier to communicate than the entire $N$ dimensional vector $\bm{\mu}$. In sub-section \ref{dist_imp}, we exploit this fact and show how this factor $\beta_i(\bm{\mu})$ may be made available to each node $i$ on-the-fly. \\
 With the stated assumptions on the cost functions, there is no duality-gap \cite{bertsekas1999nonlinear}. Convex duality theory  guarantees the existence of an optimal dual variable $\bm{\mu}^*\geq \bm{0}$ such that, the solution to the relaxed problem \eqref{dual_sep}, corresponding to $\bm{\mu}^*$, gives an optimal solution to the original problem $\textbf{P}_1$. To obtain the optimal dual variable $\bm{\mu}^*$, we solve dual $\textbf{P}_1^*$ of the problem $\textbf{P}_1$, given as follows \\\\
 \textbf{Problem} $\textbf{P}_1^*$:
\begin{framed}
\begin{eqnarray}
\textbf{Maximize} \hspace{10pt} D(\bm{\mu}) \label{dual_prob}\\
\hspace{10pt}\bm{\mu} \geq \bm{0} \nonumber
\end{eqnarray}
\end{framed}

The dual problem $\textbf{P}_1^*$ is well-known to be convex \cite{bertsekas1999nonlinear}. To solve the problem $\textbf{P}_1^*$, we use the \textbf{Projected Super-gradient} algorithm \cite{nedic2008convex}, which will be shown to be amenable to a distributed implementation.\\
At the $k$\textsuperscript{th} step of the iteration, a super-gradient $\bm{g}(\bm{\mu}(k))$ of the dual function $D(\bm{\mu})$ at the point $\bm{\mu}=\bm{\mu}(k)$ is  given by $\partial D(\bm{\mu}(k))=\bm{S}^{\text{obs}}(k)-\bm{S}(k)$ \cite{bertsekas1999nonlinear}, where $S_i^{\text{obs}}(k)$ is the observed rate of arrival of incoming user-load at the proxy of node $i$, i.e., 
 \begin{eqnarray}
 S_i^{\text{obs}}(k)\equiv \sum_{j=1}^{N}C_{ji}A_jx_j^*(k),
 \end{eqnarray}
 and $x_i^*(k)$ and $S_i^*(k)$ are the primal variables obtained from Eqn. \eqref{dual_sep}, evaluated at the current dual variable $\bm{\mu}=\bm{\mu}(k)$. Following a projected super-gradient step, the dual variables $\bm{\mu}(k)$ are iteratively updated component-wise at each node $i$ as follows: 
\begin{eqnarray} \label{sub_grad_step_1}
\mu_i(k+1) = \bigg(\mu_i(k)+ \alpha \big(S_i^{\text{obs}}(k)-S_i^*(k)\big)\bigg)^+
\end{eqnarray}
Here $\alpha$ is a small positive step-size constant, whose appropriate value will be given in Theorem \eqref{bertsekas_opt}. Since the problem-parameters might vary slowly over time, a stationary algorithm is practically preferable. Hence, we chose to use a constant step-size $\alpha$, rather than a sequence of diminishing step-sizes $\{\alpha_k\}_{k\geq 1}$.  The above constitutes theoretical underpinning of steps (5) and (6) of the proposed distributed load-management algorithm in CDN, described in Algorithm \ref{algo_dual}. \\
\subsection{Convergence of the Dual Algorithm}
To prove the convergence of the above algorithm, we first \emph{uniformly} bound the $\ell_2$ norm of the super-gradients $\bm{g}(\bm{\mu}(k))$:
\begin{eqnarray*}
\bm{g}(\bm{\mu}(k))\equiv \bm{S}^{\text{obs}}(k)-\bm{S}(k)=\bigg(\sum_{j=1}^{N}C_{ji}A_jx_j^*(k)-S_i^*(k)\bigg)_{i=1}^{N}
\end{eqnarray*} 
We start with the following lemma. 
\begin{framed}
\begin{lemma} \label{subgrad_bounded}
If the total external DNS-request arrival rate to the entire system is bounded by $A_{\max}$ (i.e. $\sum_i A_i \leq A_{\max}$) and the maximum processing-capacity of individual proxies is bounded by $T_{\max}$ (i.e. $T_i \leq T_{\max}, \forall i$) then, for all $k\geq 1$
\begin{eqnarray}
||\bm{g}(\bm{\mu}(k))||_2^2\leq A_{\max}^2 + NT_{\max}^2
\end{eqnarray}
\end{lemma} 
\end{framed}
\begin{proof}
See Appendix \ref{pf_sg_bd}. 
\end{proof}

Upon bounding the super-gradients uniformly for all $k$, the convergence of the dual algorithm follows directly from Proposition 2.2.2 and 2.2.3 of \cite{bertsekas2015convex}. In particular, we have the following theorem:
\begin{framed}
\begin{theorem} \label{bertsekas_opt}
For a given $\epsilon>0$, let the step-size $\alpha$ in Eqn. \eqref{sub_grad_step_1} be chosen as 
$\alpha = \frac{2\epsilon}{A_{\max}^2+NT^2_{\max}} $. 
Then, 
 \begin{itemize}
 \item The sequence of solutions, produced by the dual algorithm described above, converge within an $\epsilon$-neighbourhood the optimal objective value of the problem $\textbf{P}_1$.
 \item The rate of convergence of the algorithm to the $\epsilon$-neighborhood of the optima after $k$-steps is given by $c/\sqrt{k}$ where $c \sim \Theta(\sqrt{N})$.
 \end{itemize}
\end{theorem}
\end{framed}
The above result states that the rate of convergence of the dual algorithm decreases roughly at the rate of $\Theta(\sqrt{N})$, where $N$ is the total number of nodes in the system. This is expected as more nodes in the system would warrant greater amount of inter-node coordination to converge to the global optimum.

\subsection{\textsc{FastControl} Packets and Distributed Implementation} \label{dist_imp}
In the previous subsection, we derived a provably optimal load-management algorithm for CDNs, which is implementable online, provided each node $i$ knows how to obtain the value of the coupling-factors $\beta_i(\bm{\mu}(k))$ at each step, in a decentralized fashion. To accomplish this, we now introduce the novel idea of \textsc{FastControl} packets.  In brief, it exploits the underlying anycast architecture of the system (Eqn.\eqref{S_eqn}) for \emph{in-network, on-the-fly computation} of the coupling factor $\beta_i(\bm{\mu}(k))$  (Eqn. \eqref{beta_eqn}) for all $i$. 
\paragraph*{General Descriptions} \textsc{FastControl} packets are special-purpose control packets (different from the regular data-packets), each belonging to any one of $N$ distinct classes, which we refer to as \emph{categories}. The category of each \textsc{FastControl} packet is encoded in its packet-header and hence it takes extra $\log(N)$ bits to encode the categories. Physically, these control packets originate from the users and are monitored by the proxies. However, the rates at which these packets are generated, are controlled by the DNS servers of the nodes $i$ and are proportional to the dual variables $\mu_i(k)$, by a mechanism detailed below.

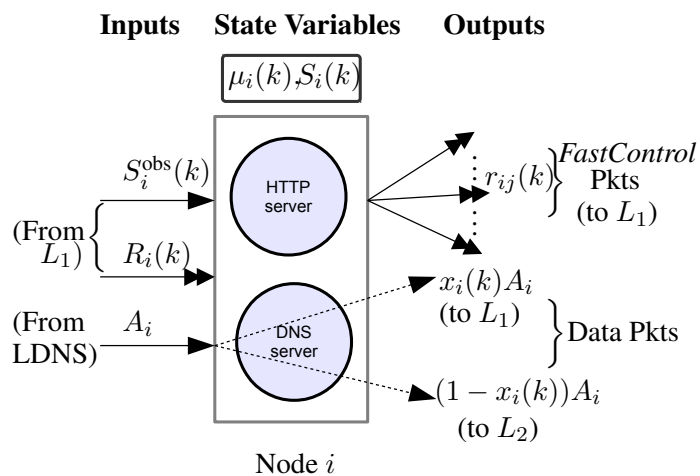
\begin{figure}
 \centering
	\begin{overpic}[scale=0.8]{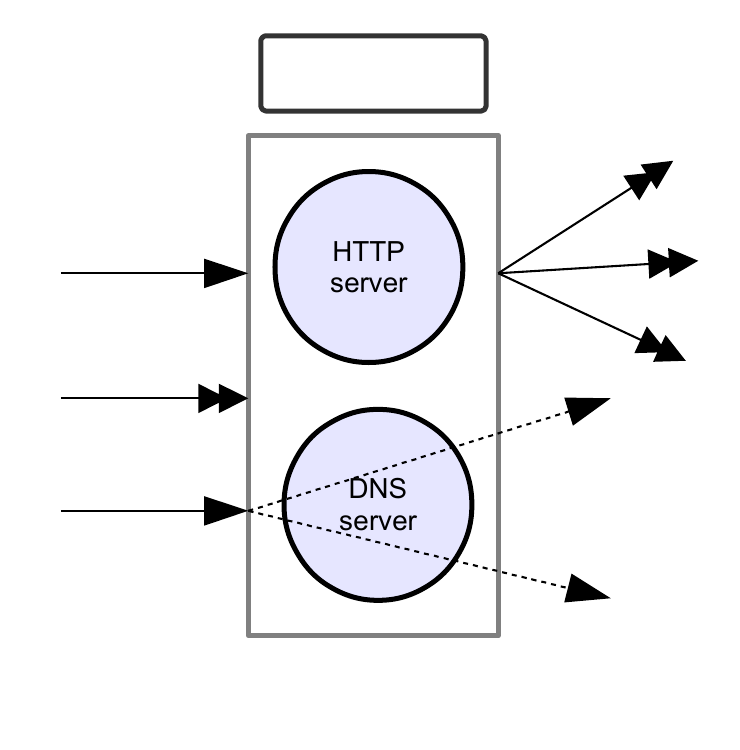}
	\put(13,68){\small{$S_i^{\text{obs}}(k)$}}
	\put(13,35){\small{$A_i$}}
	\put(-11,55){\small{(From} }
	\put(-5,50){\small{$L_1)$}}
	\put(5,54){\small{$\bigg\{$}}
	\put(13,50){\small{$R_i(k)$}}
	\put(-11,35){\small{(From} }
	\put(-11,28){\small{LDNS)}}
	\put(36,89){\small{$\mu_i(k),$}}
	\put(51,89){\small{$S_i(k)$}}
	\put(8,100){\small{\textbf{Inputs}}}
	\put(33,100){\small{\textbf{State Variables}}}
	\put(83,100){\small{\textbf{Outputs}}}
	\put(108,72){\small{\emph{FastControl}}}
	\put(115,66){\small{Pkts}}
	\put(112,59){\small{(to $L_1$)}}
	\put(89,68){\small{$\vdots$}}
	\put(89,56){\small{$\vdots$}}
	\put(92,67){\small{$r_{ij}(k)$}}
	\put(110,33){\small{Data Pkts}}
	\put(82,44){\small{$x_i(k)A_i$}}
	\put(82,37){\small{(to $L_1$)}}
	\put(81,20){\small{$(1-x_i(k))A_i$}}
	\put(86,12){\small{(to $L_2$)}}
	\put(105,66){\small{$\bigg\}$}}
	\put(105,33){\small{\bigg\}}}
	\put(42,4){\small{Node $i$}}
	\end{overpic}
\caption{\small{Node-$i$ implementing the dual algorithm}}
\label{dual_fig}
\end{figure}

\paragraph*{ Generation of FastControl packets} These packets are generated in a controlled manner by using a javascript embedded in responses to user DNS-requests (similar to how data was generated to calculate the $C$ matrix offline \cite{NSDI_paper}).  The javascript forces users to download a small image from an URL that is not affected by the load management algorithm. 
 DNS-servers in each node are configured to respond back with anycast IP address for the primary layer (i.e. $\mathcal{I}_1$) for this special DNS-query.  The use of various categories of FastControl packets will be clear from the description of the following distributed protocol used for determining $\beta_i(\bm{\mu}(k))$:\\
\begin{itemize} 
\item  At step $k$, the DNS server of each node $i$ \textbf{forces} generation of \textsc{FastControl} packets of category $j$ (through its response to DNS-queries) at the rate 
\begin{eqnarray} \label{rate_gen}
r_{ij}(k)=\gamma\mu_i(k) \frac{C_{ji}}{C_{ij}}, \hspace{10pt} j=1,2,\ldots, N
\end{eqnarray}
Note that this step is locally implementable at each node, since the value of the dual variable $\mu_i(k)$ is locally available at the node $i$. Here $\gamma>0$ is a fixed system parameter, indicating the rate of control packet generation. 
\item At each step $k$, each node $i$ also \textbf{monitors} the rate of reception of \textsc{FastControl} packets of category $i$, denoted by $R_i(k)$, at its co-located proxy. Using equation \eqref{rate_gen}, the total rate of reception $R_i(k)$ of $i$\textsuperscript{th} category \textsc{FastControl} packets at node $i$ is obtained as follows  
\begin{eqnarray*}
R_i(k)&=&\sum_{j=1}^{N}r_{ji}(k)C_{ji}= \sum_{j=1}^{N} \gamma \mu_j(k) \frac{C_{ij}}{C_{ji}} C_{ji}\\
&=& \gamma \beta_i(\bm{\mu}(k))
\end{eqnarray*}
Thus, 
\begin{eqnarray}
\beta_i(\bm{\mu}(k))= \frac{1}{\gamma}R_i(k)
\end{eqnarray}
\end{itemize}
Hence, the value of $\beta_i(\bm{\mu}(k))$ at node $i$ can be obtained locally by monitoring the rate of receptions of \textsc{FastControl} packets at the co-located proxy.
  A complete pseudocode of the algorithm is provided below. See Figure (\ref{dual_fig}) for a schematic diagram of a node implementing the algorithm. 
\begin{algorithm}[H]
 \small
    \begin{algorithmic}[1]
   
      \STATE \emph{Initialize:}  $\mu_i(0)\gets 0$
      \FOR {$k=1,2,3,\ldots $}
         \STATE \textbf{Monitor} $A_i(k), S^{\text{obs}}_i(k), R_i(k)$;
         \STATE Set $\beta_i(k)\gets \frac{1}{\gamma}R_i(k) ;$
         \STATE \emph{Update Primal variables:}
          \begin{eqnarray*}
           S_i(k)&\gets&\inf_{0\leq S_i\leq T_i} \big(g_i(S_i)-\mu_i(k)S_i\big)\\
           x_i(k)&\gets &\inf_{0\leq x_i\leq 1}\big(h_i(x_i)+A_i(k)\beta_i(k)x_i\big)
          \end{eqnarray*}
          \STATE \emph{Update Dual variable:}
          \begin{eqnarray*}
           \mu_i(k+1) \gets \big(\mu_i(k)+ \alpha (S_i^{\text{obs}}(k)-S_i(k))\big)^+
          \end{eqnarray*}

            \STATE Via DNS-response, force users to \textbf{generate} \textsc{FastControl} packets of category $j$, destined to $L_1$, at the rate
            \begin{eqnarray*} \label{generation}
             r_{ij}(k+1)= \gamma\mu_i(k+1) \frac{C_{ji}}{C_{ij}}, \hspace{10pt} \forall j=1,2,\ldots, N
            \end{eqnarray*}
         \STATE  For an incoming DNS-query, \textbf{respond} with the anycast IP-address for $L_1$ with probability $x_i(k)$ and IP-address for $L_2$ with probability $(1-x_i(k))$.
      \ENDFOR
    \end{algorithmic}
    \caption{Distributed Dual Decomposition Algorithm Running at Node $i$}
    \label{algo_dual}
    \end{algorithm}

We make the following observations :
\begin{itemize}
\item The full knowledge of the matrix $\bm{C}$ at node $i$ is also not necessary. It is sufficient that node $i$ knows the $i$\textsuperscript{th} row and column of the matrix $\bm{C}$.
\item The optimality of the algorithm \textbf{does not depend} on the diagonally dominance property of the correlation matrix $\bm{C}$, an essential requirement for the greedy heuristic (discussed in the next section) to work reasonably well.
\item The parameter $\gamma$ in Eqn. \eqref{rate_gen} is directly related to the amount of control-overhead required for the distributed algorithm. 
\end{itemize}

This completes the description of the proposed load-management algorithm, which is completely distributed and provably optimal. In the following section, we concentrate our attention on the load-management heuristic used in FastRoute and evaluate its performance. We will numerically compare the relative benefits of these two algorithms in Section \ref{simulations}. 
\section{The Greedy Load-Management Heuristic} \label{heuristic}
In this section, we focus exclusively on the distributed load management heuristic implemented in FastRoute \cite{NSDI_paper}, a commercial layered-CDN. This heuristic ignores inter-node correlations altogether. Thus, when an individual proxy becomes overloaded, the co-located DNS-server modifies its DNS-response to redirect more traffic to the data-center ($L_2$) and vice versa. This simple mechanism is reported to work well in practice when there is high correlation (60-80\%) between the node receiving the DNS-query and the node receiving the corresponding user-request. However, in the event of sudden bursts of traffic, e.g., \emph{Flash Crowds}, this greedy heuristic often leads to an uncontrollable overload situation which necessitates manual intervention \cite{NSDI_paper}. 
%
%
%

\textbf{Algorithmic challenges:} With only local information available at each DNS-server, it faces the following dilemma: offload too little to $L_2$ and the collocated proxy, if overloaded, remains overloaded; offload too much and the users are directed to remote data-centers and receive an unnecessarily delayed response, due to high round-trip latency. The coupling among the nodes because of inter-node correlation makes this problem highly challenging and gives rise to the so-called uncontrollable overload, discussed earlier in section \ref{u_o}. 



 A pseudo-code showing the general structure of FastRoute's heuristic, running at a node $i$, is provided below. An explicit control-law modeling the heuristic will be given in section \ref{analysis_greedy}. 
%

\begin{algorithm} 
\small
    \begin{algorithmic}[1]
\FOR{ $t=1,2,3,\ldots$ }
\IF{ the $i$\textsuperscript{th} proxy is \textbf{under loaded} ($S_i(t) \leq T_i$) }
\STATE \indent \indent \textbf{increase} $x_i(t)$  proportional to $-(S_i(t)-T_i)$
\ELSE 
\STATE \indent \indent \textbf{decrease} $x_i(t)$ proportional to ($S_i(t)-T_i$)
\ENDIF
\ENDFOR
 \caption{Decentralized greedy load-management heuristic used in \emph{FastRoute} \cite{NSDI_paper}, running at the node $i$}
 \label{algo_greedy}
\end{algorithmic}
\end{algorithm}

\emph{Motivation for analyzing the heuristic:} The optimal algorithm of section-\ref{optimization} requires the knowledge of the correlation matrix $\bm{C}$ and needs to utilize additional \textsc{FastControl} packets for its operation. In this section we analyze performance of the greedy heuristic, currently implemented in \emph{FastRoute} \cite{NSDI_paper}, which does not have these implementation complexities. Since this heuristic completely ignores the inter-node correlations (given by the matrix $\bm{C}$), it cannot be expected to achieve the optima of the problem $\bm{P}_1$, in general. Instead, we measure its performance by a coarser performance-metric, given by the number of proxies that undergo uncontrollable overload condition (refer to section \ref{u_o}) under its action. Depending on target applications, this metric is often practically good enough for gauging the performance of CDNs.

 \subsection{Analysis of the Greedy Heuristic} \label{analysis_greedy}
 As before, let $x_i(t)$ denote the probability that an incoming DNS-query to the node $i$ at time $t$ is returned with the anycast address of the primary layer $L_1$. Hence, the total rate of incoming load to the proxy $i$ at time $t$ is given by,
\begin{eqnarray} \label{lin}
S_i(t)=\sum_{j=1}^{N} C_{ji}A_j x_j(t) , \hspace{10pt} i=1,2,3,\ldots,N
\end{eqnarray}
The above system of linear equations can be compactly written as follows
\begin{eqnarray}
\bm{S}(t)=\bm{B}\bm{x}(t)
\end{eqnarray}
Where, 
\begin{eqnarray}
\bm{B}\equiv \bm{C}^T\textbf{diag}(\bm{A}).
\end{eqnarray} 

Let the vector $\bm{T}$ denote the processing-capacities (thresholds) of the proxies. As described above, FastRoute's greedy heuristic \eqref{algo_greedy} monitors the overload-metric $S_i(t)-T_i$ and if it is positive (i.e., the node $i$ overloaded), it reduces $x_i(t)$ (i.e., $\frac{dx_i(t)}{dt}<0$) and if the overload-metric is negative (i.e., the node $i$ under-loaded), it increases $x_i(t)$ (i.e., $\frac{dx_i(t)}{dt}>0$) proportional to the overload. We consider the following explicit control-law complying with the above general principle:
\begin{eqnarray} \label{ode}
\frac{dx_i(t)}{dt} = -\beta R(x_i(t))\big(\bm{B}\bm{x}(t)-\bm{T}\big)_i, \hspace{10pt} \forall i
\end{eqnarray}
The factor $R(x_i(t))\equiv x_i(t)(1-x_i(t))$ is a non-negative \emph{damping} component, having the property that $R(0)=R(1)=0$. This non-linear factor is responsible for restricting the trajectory of $\bm{x}(t)$ to the $N$-dimensional unit hypercube $\bm{0}\leq \bm{x}(t) \leq \bm{1}$, ensuring the feasibility of the control \eqref{ode}\footnote{Remember that $x_i(t)$'s, being probabilities, must satisfy $0\leq x_i(t)\leq 1, \forall t, \forall i$}. The scalar $\beta>0$ is a sensitivity parameter, relating the robustness of the control-strategy to the local observations at the nodes.   \\
The following theorem establishes soundness of the control \eqref{ode}:
\begin{framed}
\begin{theorem} \label{existence_uniqueness_thm}
Consider the following system of ODE
\begin{eqnarray} \label{eq1}
\dot{x_i}(t)= -R(x_i(t))(\bm{B}\bm{x}(t)-T)_i, \hspace{10pt} \forall i
\end{eqnarray}
where $R:[0,1]\to \mathbb{R}_+$ is any $\mathcal{C}^{1}$ function, satisfying  $R(0)=R(1)=0$. \\
Let $\bm{x}(0)\in \text{int}({\mathcal{H}})$, where $\mathcal{H}$ is the $N$-dimensional unit hypercube $[0,1]^{N}$. Then the system \eqref{eq1} admits a unique solution $\bm{x}(t) \in \mathcal{C}^1$ such that $\bm{x}(t) \in \mathcal{H}, \forall t\geq 0$.
\end{theorem}
\end{framed}
\begin{proof}
 See Appendix \ref{existence_uniqueness_proof}.
\end{proof}
The following theorem reveals an interesting feature of the greedy algorithm, which states that, along any periodic trajectory the average load at any node $i$ is equal to the threshold $T_i$ of that node, and hence they are stable \emph{on the average}. 
\begin{framed}
\begin{theorem} \label{limit_cycle}
Consider the system \eqref{ode} with possibly time-varying arrival rate vector $\bm{A}(t)$ such that the system operates in a periodic orbit. Then the time-averaged user-load on any node $i$ is equal to the threshold $T_i$ of that node, i.e.
\begin{eqnarray}
\frac{1}{\tau} \int_{0}^{\tau}S_i(t) dt= \bar{T}, \hspace{5pt} \forall i,
\end{eqnarray}  
where the integral is taken along a periodic orbit of period $\tau$.
\end{theorem} 
\end{framed}
\begin{proof}
See Appendix \ref{limit_cycle_proof}. 
\end{proof}

\subsection{Avoiding Locally Uncontrollable Overload}
Having established the feasibility and soundness of the control-law \eqref{ode}, we return to the original problem of locally uncontrollable overload, described in Section \ref{u_o}. In the following, we derive sufficient conditions for the correlation matrix $\bm{C}$ and the DNS-request arrival rate vector $\bm{A}$, for which the system is stable, in the sense that no locally uncontrollable overload situation takes place. 

\subsection*{Characterization of the Stability Region}
For a fixed correlation matrix $\bm{C}$, we show that if the arrival rate vector $\bm{A}$ lies within a certain polytope $\bm{\Pi}_{\bm{C}}$, the system is stable in the above sense, under the action of the greedy load-management heuristic. The formal description and derivation of the result is provided in Appendix \ref{stability_region}, which involves linearization of the ODE \eqref{ode} around certain fixed points. Here we outline a simple and intuitive derivation of the stability region $\bm{\Pi}_{\bm{C}}$. \\
We proceed by contradiction. Suppose that node $i$ is facing a locally uncontrollable overload at time $t$. Hence, by definition, the following two conditions must be satisfied at node $i$
\begin{eqnarray}
 S_i(\infty)-T_i >0 \label{ol}, \hspace{5pt}\text{and} \hspace{5pt}
 x_i(\infty)=0 \label{uc}
\end{eqnarray}
Here Eqn. \eqref{ol} denotes the fact that FastRoute node $i$ is \emph{overloaded}, i.e., the incoming traffic to node $i$'s proxy is more than the capacity of the node $i$. Eqn. \eqref{uc} denotes the fact that this overload is \emph{locally uncontrollable}, since even after node $i$'s DNS-server has offloaded \emph{all} incoming DNS-influenced arrivals to $L_2$ (the best that it can do with its local information), it is facing the overload situation. The above two equations imply that the following condition holds at the node $i$:
\begin{eqnarray} \label{cond1}
 \sum_{j \neq i} C_{ji}A_jx_j(\infty) > T_i,
\end{eqnarray}
where we have used Eqn. \eqref{lin} and the fact that $x_i(\infty)=0$. Since $0\leq x_j(\infty) \leq 1 $,
a necessary condition for uncontrollable overload \eqref{cond1} at node $i$ is $\sum_{j \neq i} C_{ji}A_j > T_i$. Thus, if $\sum_{j \neq i} C_{ji}A_j \leq  T_i$, then the locally uncontrollable overload is avoided at the node $i$ by the greedy heuristic. Taking into account all nodes, we see that if the external DNS-query arrival rate $\bm{A}$ lies in the polytope $\bm{\Pi}_{\bm{C}}$ defined as  
\begin{eqnarray} \label{polytope}
 \bm{\Pi}_{\bm{C}}=\{\bm{A}\geq \bm{0}: \sum_{j \neq i} C_{ji}A_j \leq  T_i, \hspace{5pt} \forall i=1,2,\ldots,N\}
\end{eqnarray}
then the locally uncontrollable overload situation is avoided at \emph{every} node and the system is stable. \\
Somewhat surprisingly, by exploiting the exact form of the control-law \eqref{ode}, we can  show that a two-node system (as depicted in Figure \ref{two-node_fig}) is able to control DNS-load $\bm{A}$ of any magnitude, under certain favorable conditions on the correlation matrix $\bm{C}$.


\subsubsection*{Special Case}[\textbf{Two-node System}]

\begin{figure}
 \begin{minipage}[b]{0.5\linewidth}
 \centering 
 \hspace*{-15pt} 
     \begin{overpic}[scale=0.37]{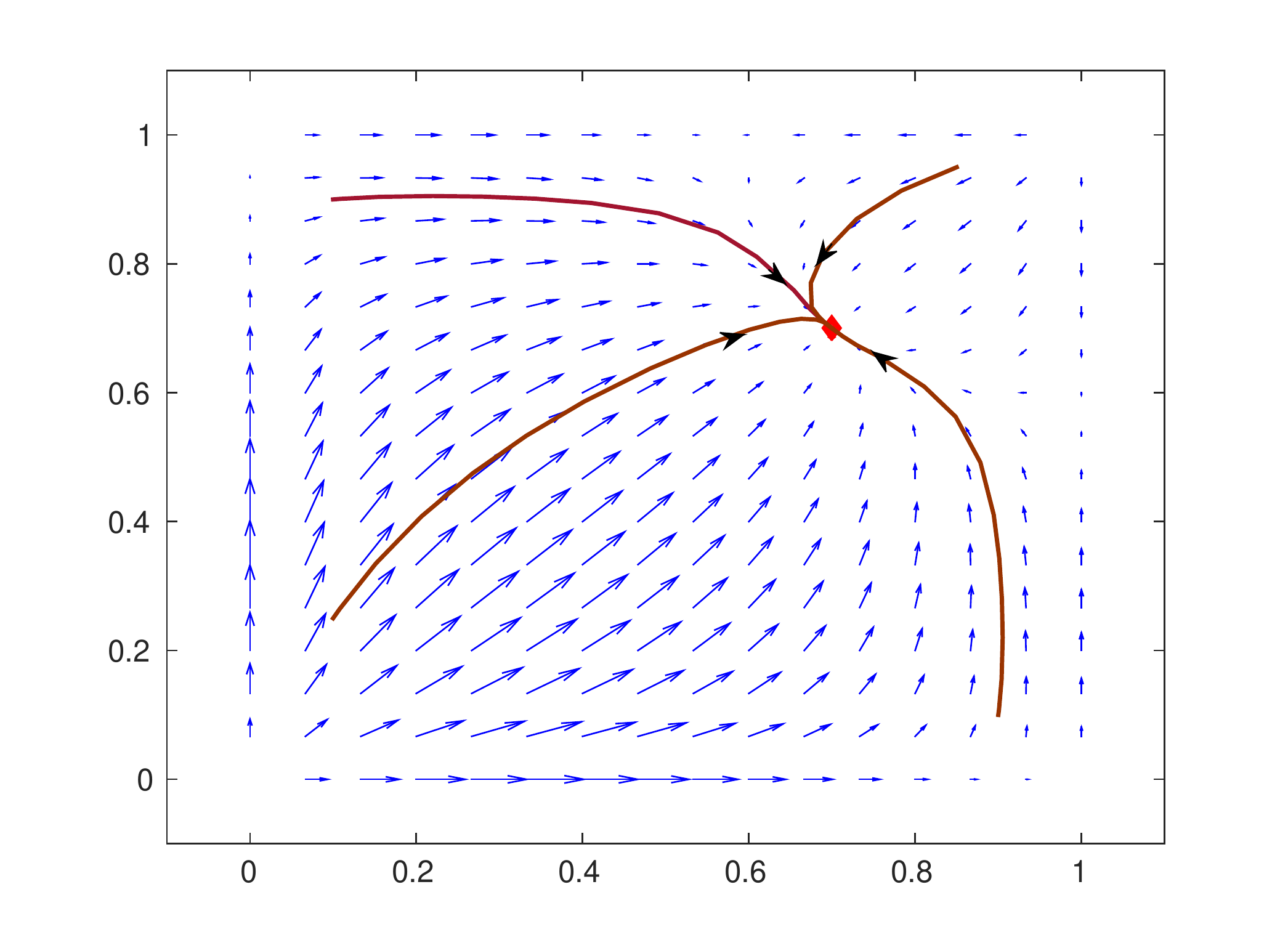}
     \put(45,-1){$\tiny{x_1(t)}$}
     \put(-0.5,38){$\tiny{x_2(t)}$}
      \put(65,49){\text{\scriptsize{(attractor)}}}
       \put(45,70){\text{\scriptsize{Fig. 4(a)}}}
    \end{overpic}
 \end{minipage}
 \begin{minipage}[b]{0.5\linewidth} 
 \centering
 \hspace*{-10pt}
  \begin{overpic}[scale=0.37]{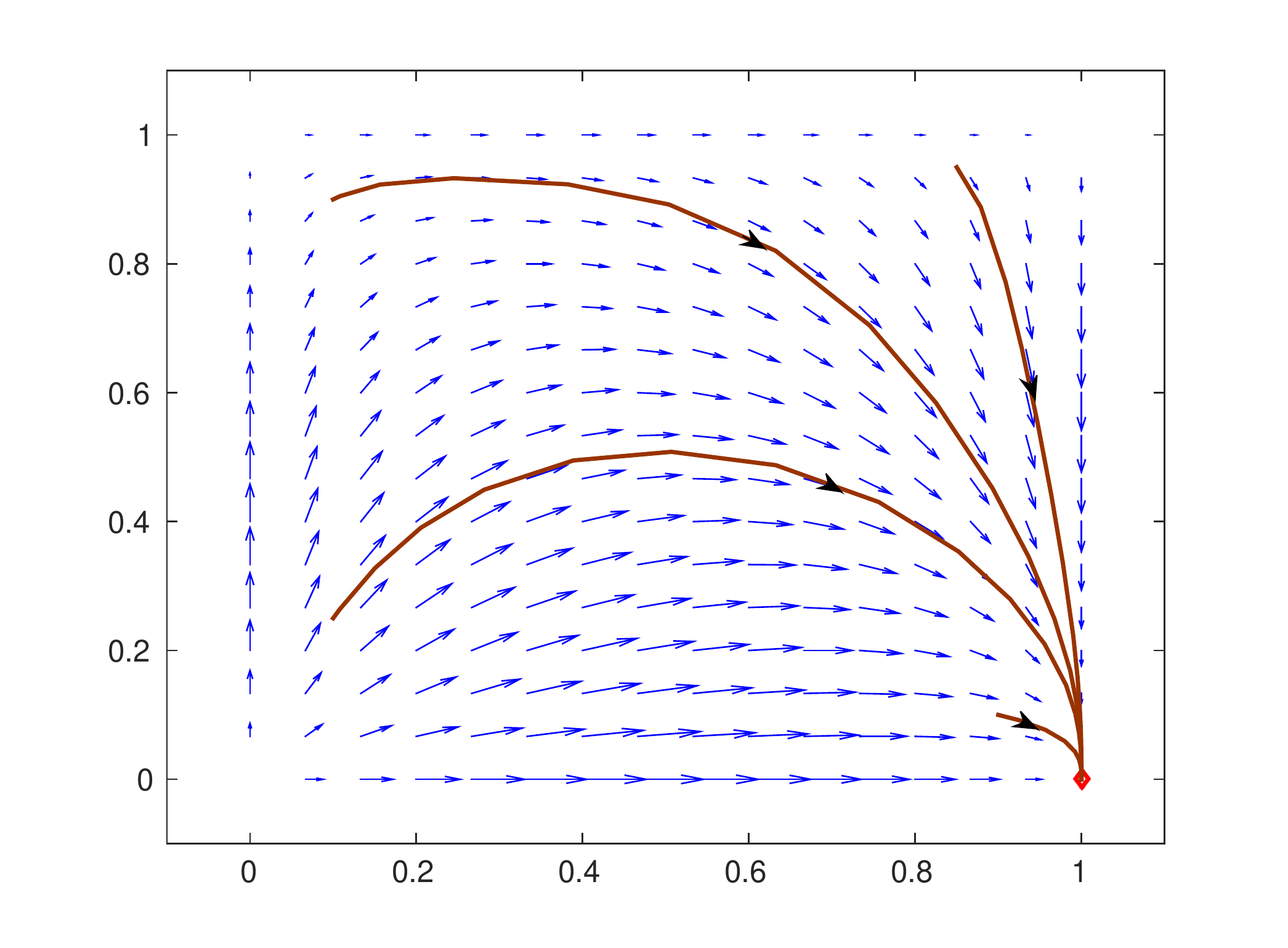}
 \put(45,-1){$\tiny{x_1(t)}$}
 \put(-0.5,38){$\tiny{x_2(t)}$}
 \put(45,70){\text{\scriptsize{Fig. 4(b)}}}
 \put(75,10){\text{\scriptsize{(attractor)}}}
   \end{overpic}
   \end{minipage}

\caption{2D vector-fields of a two-node system illustrating locally controllable (Fig. $4(a)$) and locally uncontrollable (Fig. $4(b)$) overloads.}
\label{vector_field_plot}
\end{figure}
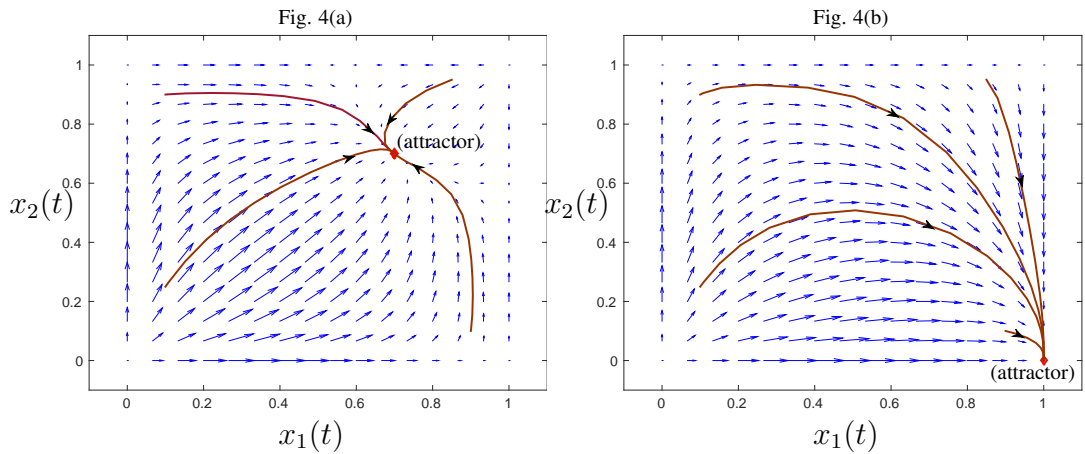

Consider a two-node CDN discussed earlier in Section \ref{optimization} (see Figure \ref{two-node_fig}). Let the correlation matrix $\bm{C}$ for the system be parametrized as follows:
\begin{eqnarray}
\bm{C}(\alpha, \beta)=
\begin{pmatrix}
\alpha && 1-\alpha \\ 1-\beta && \beta 
\end{pmatrix}
\end{eqnarray}
Then we have the following theorem :
\begin{framed}
\begin{theorem} \label{two_node_lemma}
1) The system does not possess \emph{any} periodic orbit for \emph{any} values of its defining parameters: $\bm{A},\bm{C}(\alpha,\beta),\bm{T}$. Thus the two-node CDN never oscillates.\\
2) If $\alpha > \frac{1}{2}$ and $\beta > \frac{1}{2}$ then the system is locally controllable (i.e., no locally uncontrollable overload) for \emph{all} arrival rate-pairs $(A_1,A_2)$.\\
3) If $\alpha < \frac{1}{2}$ and $\beta < \frac{1}{2}$ then a sufficient condition for local controllability of the system is $A_1<\frac{T_1}{1-\alpha}$, $A_2 < \frac{T_2}{1-\beta}$. 
\end{theorem}
\end{framed}
\begin{proof}
The proof of part-(1) follows from Dulac's criterion \cite{strogatz2014nonlinear}, whereas proof for part-(2) and (3) follows from linearization arguments. See Appendix \ref{two_node_lemma_proof} for details.
\end{proof}
We emphasize that the part-(2) of the Theorem \ref{two_node_lemma} is surprising, as it shows that the system remains locally controllable, no matter how large the incoming DNS-request arrival rate be (c.f. Section \ref{u_o}) . The 2D vector-field plot in Figure \ref{vector_field_plot} illustrates the above result. In Figure 4(a), the matrix $\bm{C}$ is taken to be one satisfying the condition of part (2) of lemma \ref{two_node_lemma}. As shown, all four phase-plane trajectories with different initial conditions converge to an interior fixed point $(x_1(\infty)>0, x_2(\infty)>0)$. \\
For the purpose of comparison, in Figure 4(b) we plot the 2D vector-field  of a locally uncontrollable system (e.g., the system in Fig. \ref{two-node_fig}). It is observed that all four previous trajectories converge to the uncontrollable attractor $(x_1(\infty)=1,x_2(\infty)=0)$. From the vector-field plot, it is also intuitively clear why a periodic orbit can not exist in the system.

\paragraph*{Application of Theorem \ref{two_node_lemma}}
Consider a setting where, instead of making locally greedy offloading decisions, nodes are permitted to partially coordinate their actions. Also assume that there exists a partition of the set of nodes into two non-empty disjoint sets $G_1$ and $G_2$, such that their effective self-correlation values $\alpha(G_1)$ and $\beta(G_2)$, defined by 
\begin{eqnarray*}
\alpha(G_1)=\frac{\sum_{i \in G_1}\sum_{j \in G_1}C_{ij}}{|G_1|}, \beta(G_2)=\frac{\sum_{i \in G_2}\sum_{j \in G_2}C_{ij}}{|G_2|}
\end{eqnarray*}
satisfy the condition (2) of Theorem \ref{two_node_lemma}, i.e. $\alpha(G_1)>\frac{1}{2}, \beta(G_2)> \frac{1}{2}$. Then if the nodes in the sets $G_1$ and $G_2$ coordinate and jointly implement the greedy policy, then the system is locally controllable for all symmetric arrivals.  




\section{Numerical Evaluations}\label{simulations}
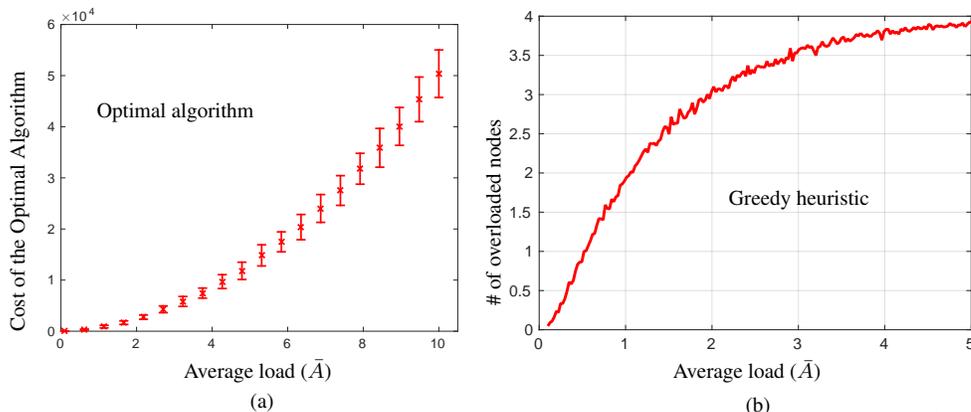
\begin{figure}
 \begin{minipage}[b]{0.45\linewidth}
 \centering
 \hspace*{-10pt} 
     \begin{overpic}[scale=0.32]{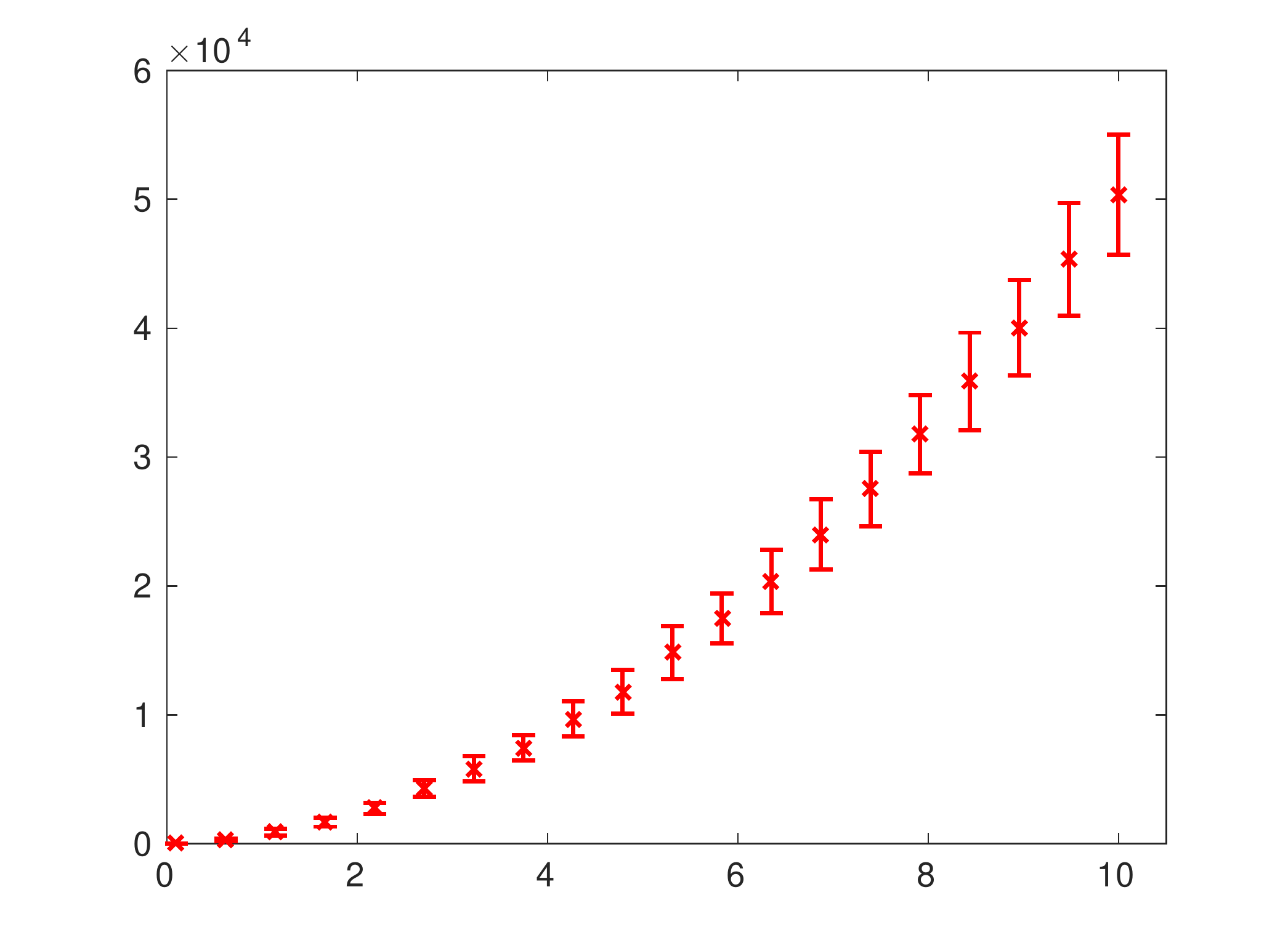}
     \put(37,-1){\scriptsize{Average load ($\bar{A}$)}}
     \put(50,-7){\scriptsize{(a)}}
    \put(20,50){\scriptsize{Optimal algorithm}}
     \put(3,8){\rotatebox{90}{\scriptsize{Cost of the Optimal Algorithm}}}
    \end{overpic}
  \end{minipage}  
   \begin{minipage}[b]{0.45\linewidth}
    \label{greedy_overload_plot}
    \centering
    \hspace*{-10pt} 
  \begin{overpic}[scale=0.27]{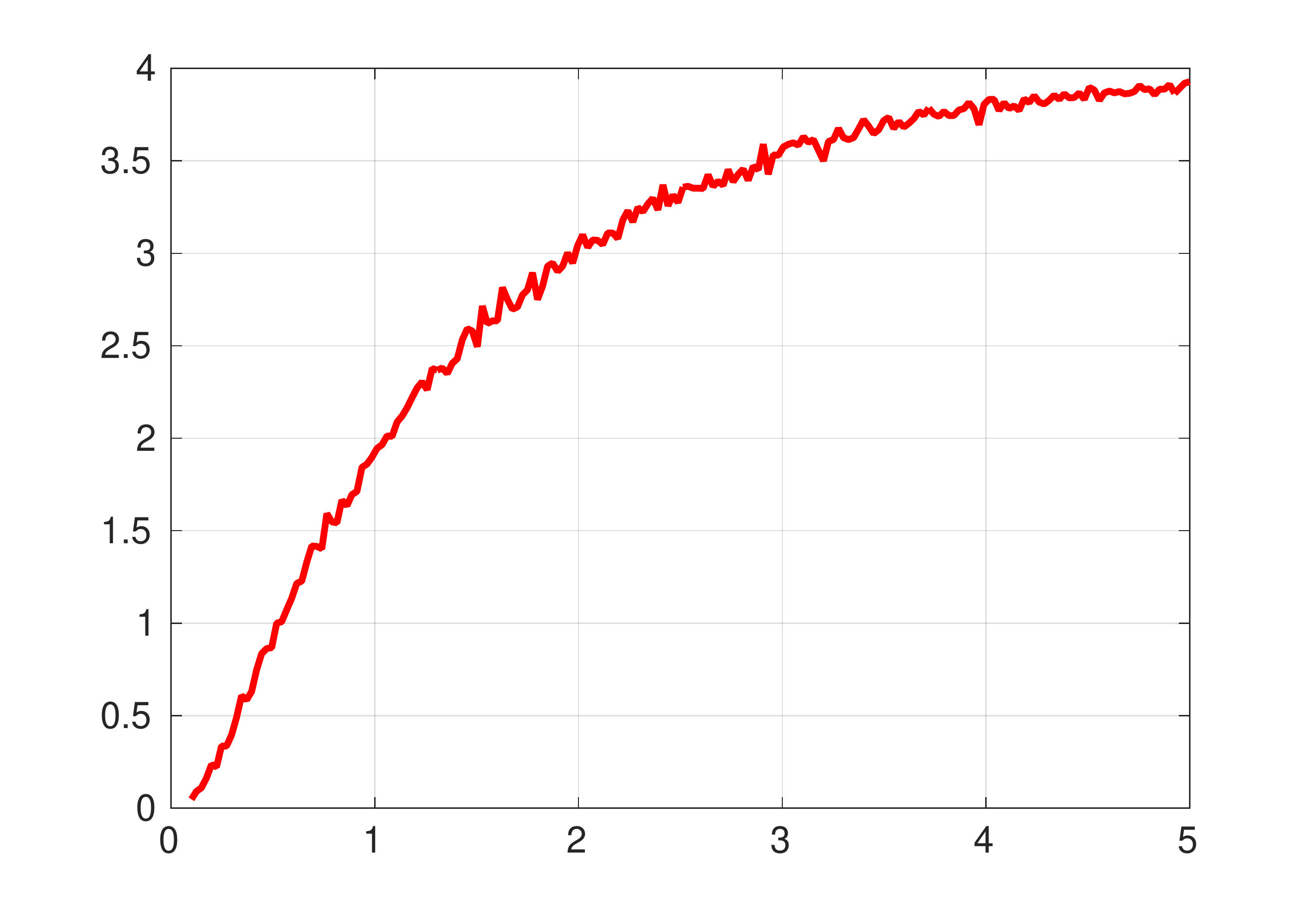}
 \put(37,-1){\scriptsize{Average load ($\bar{A}$)}}
 \put(47,30){\scriptsize{Greedy heuristic}}
 \put(50,-7){\scriptsize{(b)}}
 \put(3.5,12){\rotatebox{90}{\scriptsize{\# of overloaded nodes}}}
   \end{overpic}
\end{minipage}
 \caption{(a) Statistical variation of the cost of the optimal algorithm with mean load (DNS-query arrival rate) $\bar{A}$. (b) Average number of nodes undergoing uncontrollable overload condition under the action of the greedy algorithm (Threshold $T_i=0.7$ for all nodes). Total number of nodes in the system included in this study is $N=60$.}
\label{cost_function_opt}
\end{figure}

We use the operational FastRoute CDN to identify relevant system parameters for critical evaluations of the optimal algorithm and the heuristic. Currently, FastRoute has many operational nodes, spreading throughout the world \cite{NSDI_paper}. We show a sample result with $N=60$ nodes. The inter-node correlation matrix $\bm{C}$ is computed using system-traces spanning over three months.\\
For our performance evaluations, we use the cost-functions given in Eqns. \eqref{ex1} and \eqref{ex2}. Different (normalized) parameters appearing in the cost-functions are chosen as follows 
\begin{eqnarray}
\gamma_i=1, \theta_i=10, \eta_i=1, T_i=0.7, d_i \sim U_i[0,1]\hspace{25pt}\forall i
\end{eqnarray}
where $U_i[0,1]$'s denote i.i.d. uniformly distributed random variables in the range $[0,1]$. The DNS-query arrival rates $A_i$'s are assumed to be i.i.d. distributed  according to a Poisson variable with mean $\bar{A}$, which varies across the range $[0.1,10]$. \\
The optimal solutions of the Lagrangian relaxations in Eqns. \eqref{dual_sep} for the above cost-functions can be obtained in closed form as follows :
\begin{eqnarray}
S_i^*(\bm{\mu})=T_i\max \bigg(0, 1-\sqrt{\frac{\eta_i}{\mu_i}}\bigg)
\end{eqnarray}
and,
\begin{eqnarray}
x_i^*(\bm{\mu})=\begin{cases}
1, \hspace{15pt}\text{if  }c_{2i}>0\\
1+\frac{c_{2i}}{2c_{1i}}, \text{  if  } c_{2i}\leq 0 \text{ and }2c_{1i}\geq -c_{2i}\\
0, \hspace{15pt}\text{o.w.}
\end{cases}
\end{eqnarray}
where $c_{1i}\equiv A_i\theta_i$ and $c_{2i}\equiv \theta_i d_i-\beta_i(\bm{\mu})$.\\
For each value of $\bar{A}$, we run the simulation $N_E=100$ times, by randomizing over both $A_i$ and $d_i, \forall i$. The mean and standard-deviation of the resulting optimal cost values are plotted in the Figure \ref{cost_function_opt}(a). As expected, the average cost increases as the overall DNS-query arrivals to the system increases. However, the resulting cost remains finite always. This implies that \emph{none} of the proxies are overloaded. \\
The above observation is in sharp contrast with the situation using the greedy-heuristic, the subject of Figure \ref{cost_function_opt}(b). Here we plot the number of overloaded proxies for different values of $\bar{A}$, keeping all other system-parameters the same. While the greedy algorithm does yield acceptable result for small values of $\bar{A}<< T_i=0.7$, we see that as many as four proxies undergo locally uncontrollable overload for relatively large values of $\bar{A}$.  \\
Thus, depending on the computed correlation-matrix $\bm{C}$ and a projected bound of the DNS-query arrival rate $\bm{A}$, we can make an informed decision about the choice of the algorithms to employ in a CDN and the inherent complexity-vs-optimality trade-off it entails. 

\section{Conclusion}
In this paper we formalize the load management problem in modern CDNs utilizing anycast. We first formulate the problem as a convex optimization problem and study its dual and an associated algorithm. The novel idea of \textsc{FastControl} packets facilitates distributed implementation of the proposed dual algorithm. Next we analyze stability properties of a greedy heuristic, currently in operation in a major commercial CDN. We find that the optimal algorithm significantly out-performs the heuristic for moderate-to-high value of system load. However, the heuristic performs satisfactorily given that the system is \emph{loosely-coupled} and the offered load is low. Thus an informed choice between these two algorithms may be made depending on the range of system-parameters and the desired optimality/complexity trade-off for a particular CDN. Future work would involve investigating the amount of \textsc{FastControl} packets necessary for the dual algorithm to work in the presence of random packet-loss and delayed feedback. Also, It would be interesting to generalize the findings of Theorem \ref{two_node_lemma} for the case of more than two nodes.

\section*{References}
\bibliographystyle{elsarticle-harv}
\bibliography{Sigcomm_submission}
\section{Appendix}
\subsection{Proof of Lemma \ref{subgrad_bounded}} \label{pf_sg_bd}
\begin{proof}
We have, 
\begin{eqnarray}
||\bm{g}_k||_2^2 &=& \sum_{i=1}^{N} (S_i^{\text{obs}}(k)-S_i(k))^2 \\
&\leq & \sum_{i=1}^{N} \big(S_i^{\text{obs}}(k)\big)^2 + \sum_{i=1}^{N}S_i^2(k) \label{non_neg}\\
&\leq & \sum_{i=1}^{N} \big(\sum_{j=1}^{N}C_{ji}A_jx_j(k)\big)^2 + NT_{\max}^2 \label{feas1}\\
&\leq & \big(\sum_{i=1}^{N}\sum_{j=1}^{N}C_{ji}A_j\big)^2 + NT_{\max}^2 \label{feas2}\\
&=& \big(\sum_{j=1}^{N}A_j \sum_{i=1}^{N}C_{ji}\big)^2 + NT_{\max}^2 \label{order} \\
& \leq & (\sum_{j=1}^{N}A_j)^2+ NT_{\max}^2 \label{corr_mat}\\
&\leq & A_{\max}^2 + NT^2_{\max}
\end{eqnarray}
Here Eqn. \eqref{non_neg} follows from non-negativity of $S_i^{\text{obs}}$ and $S_i$, Eqn. \eqref{feas1} follows from the defining equation of $S_i^{\text{obs}}$ and the constraint that $S_i \leq T_i$ (viz. Eqn. \eqref{dual_sep}), Eqn. \eqref{feas2} follows from the constraint $0\leq x_i \leq 1, \forall i$, Eqn. \eqref{order} follows from the change of order of summation and finally Eqn. \eqref{corr_mat} follows from the fact that $\bm{C}$ is a correlation matrix and hence its rows sum to unity (viz. Eqn. \eqref{corr_mat_prop}). 
\end{proof}
\subsection{Proof of Theorem \ref{existence_uniqueness_thm}}
\label{existence_uniqueness_proof}
\begin{proof} 
First we show that, any solution of the system \eqref{ode} (if exists) must lie in the unit hypercube $\mathcal{H}$. We prove it via contradiction. On the contrary to the claim, assume that for some solution of the system \eqref{ode}, there
 exists a component $x_i(\cdot)$ and a finite time $0\leq \tau < \infty$ such that $x_i(\tau) <0$. Since $x_i(\cdot)$ is continuous and $x_i(0)>0$, by \emph{intermidiate value theorem}, there must exist a time $0<t_0<\tau$ such that $x_i(t_0)=0$. Now consider the differential equation corresponding to the $i$\textsuperscript{th} component of the system \eqref{eq1}. We substitute for all other components $\{x_j(t), j\neq i\}$ on the RHS of the following equation.
 
\begin{eqnarray} \label{eq2}
 \dot{x_i}(t)= -R(x_i)(\sum_{j}B_{ij}x_j(t)-T_i), \hspace{20pt} x_i(t_0)=0
\end{eqnarray}
 Since the vector $\bm{x}(t)$ is $\mathcal{C}^{1}$ and the regularizer $R(\cdot)$ is assumed to be $\mathcal{C}^{1}$, the RHS of the equation \eqref{eq2} is $\mathcal{C}^1$. Hence \eqref{eq2}  admits a \emph{unique local solution}. However note that the following is a solution to \eqref{eq2}
 \begin{eqnarray} \label{eq3}
  x_i(t)=0, \hspace{20pt} \forall t\geq t_0
 \end{eqnarray}
 This is because $R(0)=0$. By uniqueness, \eqref{eq3} is the \emph{only} solution to \eqref{eq2}. This contradicts the fact that $x_i(\tau)<0$. Hence $\bm{x}(t) \geq \bm{0}, \forall t\geq 0$. In a similar fashion, we can also prove that $\bm{x}(t) \leq \bm{1}, \forall t \geq 0$. This proves that all solutions to \eqref{ode} must lie in the compact set $\mathcal{H}$. \\
 To complete the proof, we observe that the RHS of  the system \eqref{ode} is locally Lipschitz at each point in the compact set $\mathcal{H}$. Thus the global existence of the solution of \eqref{ode} follows directly from Theorem 2.4 of \cite{khalil1996nonlinear}. 
 
 \end{proof}
 
 \subsection{Proof of Theorem \ref{limit_cycle}} \label{limit_cycle_proof}
 \begin{proof}
  Let $\tau$ be the period of the orbit. Consider the $i$\textsuperscript{th} differential equation 
   \begin{eqnarray} 
\dot{x_i}(t)= -R(x_i)(S_i(t)-T_i)\\
\frac{dx_i}{R(x_i)}= - (S_i(t)-T_i)dt
 \end{eqnarray}
 Since $x_i(\cdot)$ belongs to the \emph{interior} of the compact set $\mathcal{H}$, $\frac{1}{R(x_i(t))}$ does not have a zero in the denominator for the enire orbit. Hence $\frac{1}{R(x_i(t))}$ is continuous and its Riemann integral exists \cite{rudin1964principles}. Integrating both sides from $0$ to $\tau$, we have 
 \begin{eqnarray} \label{eq4}
  \int_{0}^{\tau} \frac{dx_i}{R(x_i)} = \int_{0}^{\tau} (S_i(t)-T_i) dt
 \end{eqnarray}
Let $J(x_i)$ be an anti-derivative of $\frac{1}{R(x_i)}$. Hence using the fundamental theorem of calculus \cite{rudin1964principles}, we can write the LHS of \eqref{eq4} as 
\begin{eqnarray}
 J(x_i(\tau))- J(x_i(0))= \int_{0}^{\tau} S_i(t)dt - \tau T_i
 \end{eqnarray} 
Since the orbit is assumed to have a period $\tau$, we have $x_i(\tau)=x_i(0)$. Hence $ J(x_i(\tau))- J(x_i(0))=0$. Thus we have
\begin{eqnarray*}
 \bar{S_i}\equiv\frac{1}{\tau}\int_{0}^{\tau}S_i(t)dt = T_i
\end{eqnarray*}
 \end{proof}
 
 \subsection{Formal Derivation of the Stability Condition} \label{stability_region}
 Let us write the system \eqref{ode} conveniently as $\dot{\bm{x}}= \bm{F}(\bm{x})$. Consider a fixed point $\overline{x}$ of the system such that the $k$\textsuperscript{th} node faces an uncontrollable overload condition. By \emph{Hartman-Grobman} theorem \cite{strogatz2014nonlinear}, it is enough to consider linearized version of the system to determine the stability of fixed points. The first-order Taylor expansion about the fixed point $\overline{x}$ yields the following:
\begin{eqnarray*}
\dot{\bm{x}}\approx F(\bar{\bm{x}}) + \frac{\partial \bm{F}(\bm{x})}{\partial{\bm{x}}}|_{\bm{x}=\overline{\bm{x}}}(\bm{x}-\bar{\bm{x}})= \frac{\partial \bm{F}(\bm{x})}{\partial{\bm{x}}}|_{\bm{x}=\overline{\bm{x}}}(\bm{x}-\bar{\bm{x}})
\end{eqnarray*} 
Where, $\frac{\partial \bm{F}(\bm{x})}{\partial{\bm{x}}}|_{\bm{x}=\overline{\bm{x}}}$ denotes the Jacobian \cite{ogata1970modern} of the system \eqref{ode} evaluated at the fixed point $\bm{x}=\overline{\bm{x}}$ and $B_{ij}=C_{ji}A_j$. The second equation follows because $\overline{\bm{x}}$ is assumed to be a fixed point (and consequently $\bm{F}(\overline{\bm{x}})=\bm{0})$. \\
Next we proceed to explicitly compute the Jacobian of the system \eqref{ode} at a given point $\bm{x}$. Note that the $i$\textsuperscript{th} row of the system equation is given by 
\begin{eqnarray}
F_i(\bm{x})\equiv -x_i(1-x_i)(\sum_{j}B_{ij}x_j - T)
\end{eqnarray}
Thus for $i \neq j$, we have 
\begin{eqnarray} \label{off_diag}
\frac{\partial F_i}{\partial x_j}=-x_i(1-x_i)B_{ij}
\end{eqnarray}
and for $i=j$, the diagonal entry is given by
\begin{eqnarray} \label{diag}
&& \frac{\partial F_i}{\partial x_i}=-\bigg(x_i(1-x_i)B_{ii} + (1-2x_i)(\sum_{j}B_{ij}x_j - T)\bigg) \nonumber \\
&=& -\bigg(x_i(1-x_i)B_{ii} + (1-2x_i)(S_i-T)\bigg)
\end{eqnarray}
Since the node $k$ is assumed to undergo an uncontrollable overload condition, we necessarily have $x_k=0$. Hence from Eqns \eqref{off_diag} and \eqref{diag}, in the $k$\textsuperscript{th} row of the Jacobian matrix, the off-diagonal entries are all zero and the diagonal entry is $-(S_k-T)$. Hence if $S_k-T<0$, at least one eigenvalue of the Jacobian matrix at the fixed point $\overline{x}$ is strictly positive and hence the fixed point $\overline{x}$ is unstable. This implies that a sufficient condition to avoid uncontrollable overload at node $k$ is given by 
\begin{eqnarray} \label{node_k_uo}
\sum_{j\neq k}C_{jk}A_j \leq T_k
\end{eqnarray}
where we have used the fact that $x_i(t)\leq 1, \forall i$ and $x_k=0$. The derivation of the sufficient condtion for absence of uncontrollable overload condition is completed by taking intersection of hyperplanes \eqref{node_k_uo} for all $k=1,2,\ldots, N. \hspace{5pt} \blacksquare$  
\subsection{Proof of Theorem \ref{two_node_lemma}} \label{two_node_lemma_proof}
\begin{proof}
\textbf{part-(1) [non-existence of periodic orbits]}\\
We use Dulac's criterion to prove the non-existence of periodic orbits for the general two-node system. For ease of reference, we recall Dulac's criterion below :
\begin{theorem}[Dulac's criterion \cite{strogatz2014nonlinear}]
Let $\dot{\bm{x}}=\bm{f}(\bm{x})$ be a continuously differentiable vector-field defined on a simply connected subset $R$ of the plane. If there exists a continuously differentiable, real-valued function $g(\bm{x})$ such that $\nabla \cdot (g \dot{\bm{x}})$ has one sign throughout $\mathcal{D}$, then there are no closed orbits lying entirely in $\mathcal{D}$.
\end{theorem}
Now we return to the proof of the result. Let $\mathcal{H}_2=[0,1]^2$ be the unit square, where by virtue of theorem \ref{existence_uniqueness_thm}, the trajectory of the two-node system lies for all time $t\geq 0$. Thus, it suffices to show that there does not exist any periodic orbit in its interior $\mathring{\mathcal{H}_2}\equiv \mathcal{D}$. It is obvious that the region $\mathcal{D}$ is simply connected. Now consider the following $g(\bm{x})$ for application of the Dulac's criterion, 
\begin{eqnarray}
g(\bm{x})=\frac{1}{x_1x_2(1-x_1)(1-x_2)}
\end{eqnarray}
It is easy to verify that $g(\bm{x})$ is continuously differentiable throughout the region $\mathcal{D}$. We next evaluate the divergence 
\begin{eqnarray}
\nabla \cdot (g\dot{\bm{x}})= -\beta\bigg(\frac{B_{11}}{x_2(1-x_2)}+\frac{B_{22}}{x_1(1-x_1)}\bigg)
\end{eqnarray} 
It is easy to verify that the term within the parenthesis is always strictly positive throughout the region $\mathcal{D}$. Hence, by Dulac's criterion, there are no closed orbits in $\mathcal{D}$. This proves the result.  \\
\textbf{part-(2) and (3) [controllability of the system]}\\
Let the arrival rates to node 1 and 2 be given by $A_1$ and $A_2$. Let $x_1$ and $x_2$ denote the operating point of the system at the steady-state. Our objective is to find sufficient conditions on the arrival rate vector $(A_1,A_2)$, under which the operating points $(x_1,x_2)$ such that either $x_1=0$ or $x_2=0$ (i.e. full offload to Layer-II) are avoided in the \emph{steady-state}. This will ensure that no uncontrollable overload situation takes place in the system. First we consider the fixed point 
\begin{eqnarray}
x_1=0, x_2=1
\end{eqnarray}
This fixed point will be stable if both the eigenvalues of the Jacobian matrix at this point be negative. From Eqns. \eqref{off_diag} and \eqref{diag} we have the following two conditions: 
\begin{eqnarray*}
S_1>T, S_2<T
\end{eqnarray*}
i.e.,
\begin{eqnarray*}
(1-\alpha)A_2 > T, \beta A_2 <T
\end{eqnarray*}
i.e., 
\begin{eqnarray} \label{st_1}
\frac{T}{1-\alpha}<A_2< \frac{T}{\beta} 
\end{eqnarray}
Similarly, analyzing the stability of the fixed points around the point $x_1=1, x_2=0$, we obtain that this fixed point will be unstable if 
\begin{eqnarray*}
S_1<T, S_2>T
\end{eqnarray*}
i.e.,
\begin{eqnarray*}
\alpha A_1 < T, (1-\beta)A_1 > T
\end{eqnarray*}
i.e.,
\begin{eqnarray} \label{st_2}
\frac{T}{1-\beta}< A_1 < \frac{T}{\alpha} 
\end{eqnarray}
Note that if $\alpha + \beta >1 $, both the above regions \eqref{st_1} and \eqref{st_2} will be empty and the fixed points $(1,0)$ and $(0,1)$ will be always unstable. Thus the operating point will not converge to these undesirable fixed points in the steady-state. \\
Now consider the (possibly feasible) fixed point $(x_1,x_2)$ such that
\begin{eqnarray}
x_1=0, S_2=T
\end{eqnarray}
The above condition translates to,
\begin{eqnarray}
x_1=0, A_2 x_2 \beta = T \nonumber \\
x_1=0, x_2=\frac{T}{A_2\beta}
\end{eqnarray}
This fixed point will be feasible provided $\frac{T}{A_2\beta}<1$. The Jacobian matrix about this fixed point is evaluated as 
\begin{eqnarray*}
\frac{\partial \bm{F}(\bm{x})}{\partial{\bm{x}}}|_{\bm{x}}=-
\begin{pmatrix}
(1-\beta)\frac{T}{\beta}-T && 0 \\
\frac{T}{A_2\beta}(1-\frac{T}{A_2\beta}) B_{21} && \frac{T}{A_2\beta}(1-\frac{T}{A_2\beta})B_{22}
\end{pmatrix}
\end{eqnarray*}
From the Jacobian matrix above, we conclude that both of its eigenvalues are negative provided the following two conditions hold

\begin{eqnarray}
A_2> \frac{T}{\beta} \hspace*{2pt}, \hspace*{2pt} \beta<\frac{1}{2}.
\end{eqnarray}
 Doing similar analysis around the fixed point $(S_1=T, x_2=0)$, we conclude that the above fixed point will be stable for all arrival rates $A_1> \frac{T}{\alpha}$ and $\alpha <\frac{1}{2}$. \\
Finally, to obtain efficient operating region for the system (with no uncontrollable overload situation), we take union over stability region of all undesired fixed points and take the complement of it. Hence, if $\alpha > \frac{1}{2},\beta > \frac{1}{2}$ all the undesrible fixed points are unstable and hence the uncontrollable overload situation is avoided for \emph{all} DNS-request arrival rates $\bm{A}$. This proves part (2) of the theorem. \\
 On the other hand, if either $\alpha < \frac{1}{2}$ or $\beta< \frac{1}{2}$ holds, then a sufficient condition for the stability of the system is given by 
\begin{eqnarray*}
A_1<\frac{T}{1-\alpha}, A_2<\frac{T}{1-\beta}
\end{eqnarray*}
This proves part (3) of the theorem. 
\end{proof}

\end{document}